\documentclass{aa}  

\usepackage{graphicx}
\usepackage{txfonts}
\usepackage{color}
\usepackage{natbib,twoopt}
\usepackage[breaklinks=true]{hyperref} 
\bibpunct{(}{)}{;}{a}{}{,}             
\makeatletter
  \newcommandtwoopt{\citeads}[3][][]{\href{http://adsabs.harvard.edu/abs/#3}%
    {\def\hyper@linkstart##1##2{}%
     \let\hyper@linkend\@empty\citealp[#1][#2]{#3}}}
  \newcommandtwoopt{\citepads}[3][][]{\href{http://adsabs.harvard.edu/abs/#3}%
    {\def\hyper@linkstart##1##2{}%
     \let\hyper@linkend\@empty\citep[#1][#2]{#3}}}
  \newcommandtwoopt{\citetads}[3][][]{\href{http://adsabs.harvard.edu/abs/#3}%
    {\def\hyper@linkstart##1##2{}%
     \let\hyper@linkend\@empty\citet[#1][#2]{#3}}}
  \newcommandtwoopt{\citeyearads}[3][][]%
    {\href{http://adsabs.harvard.edu/abs/#3}
    {\def\hyper@linkstart##1##2{}%
     \let\hyper@linkend\@empty\citeyear[#1][#2]{#3}}}
\makeatother

\begin{document}

   \title{Study of an active region prominence using spectropolarimetric data in the \ion{He}{i} D$\mathrm{_{3}}$ multiplet}
   
   \author{S. Esteban Pozuelo
   \inst{1,2}
   \and
   A. Asensio Ramos
   \inst{1,2}
   \and
   J. Trujillo Bueno
   \inst{1,2,3}
   \and
   R. Ramelli
   \inst{4}
   \and
   F. Zeuner
   \inst{4}
   \and
   M. Bianda
   \inst{4}
   }
	
   \institute{Instituto de Astrof\'{i}sica de Canarias, C/ V\'{i}a L\'{a}ctea, s/n, 38205 La Laguna, Tenerife, Spain\\
              \email{sesteban@iac.es}
         \and
             Departamento de Astrof\'{i}sica, Universidad de La Laguna, 38205 La Laguna, Tenerife, Spain
         \and
        	Consejo Superior de Investigaciones Científicas, Spain   
         \and
         Istituto Ricerche Solari (IRSOL), Università della Svizera italiana (USI), CH-6605 Locarno-Monti, Switzerland
        }
             
\authorrunning{Esteban Pozuelo et al.}
   \date{Received ; accepted}

 
  \abstract
   {Prominences are cool overdensities of plasma supported by magnetic fields that levitate in the solar corona. The physical characterization of these structures is key for understanding the magnetic field in the corona.}
   {Our work attempts to shed light on the properties of prominences by using observations at high polarimetric sensitivity in the \ion{He}{i}~D$\mathrm{_{3}}$ multiplet taken with the Z\"urich Imaging Polarimeter-3 instrument at the Istituto ricerche solari Aldo e Cele Dacc\`o observatory.}
   {We used the Hanle and Zeeman light inversion code to infer the thermodynamic and magnetic properties of an active region prominence, assuming one- and two-component models.}
   {Our observations unveil a great diversity of physical conditions in the prominence. The observed Stokes profiles are usually broad and show interesting features, which can be described assuming a two-component model. The contribution of each component and the trends inferred for some parameters vary with the distance to the solar limb. While both components have analogous properties and contribute similarly close to the limb, a major component mainly describes the properties inferred at 10--40\arcsec \,away from the limb. Moreover, both components usually show significant differences in thermal broadening, which is essential for ensuring a good fit quality between observations and synthetic profiles. Summarizing, the observed region of the prominence shows line-of-sight velocities of 1--3~km~s$\mathrm{^{-1}}$ and rather horizontal fields of 20--80~gauss. We also report hints of a twist close to a prominence foot and changes in the magnetic configuration at specific locations.}
   {Our results indicate a mainly horizontal magnetic field of a few tens of gauss in the prominence. A model of two components with different thermal broadenings and filling factors, depending on the limb distance, is crucial for providing a consistent solution across most of the observed prominence.} 


   \maketitle
%

\section{Introduction}
\label{sec:intro}

The solar limb exhibits a myriad of phenomena in observations acquired in chromospheric and transition region lines. Among these phenomena, there are bright clouds of plasma that manifest in the shape of arcades, called prominences. Since long ago, prominences have attracted much attention and great efforts have been made to characterize them \citepads[e.g.,][]{2010SSRv..151..243L, 2010SSRv..151..333M, 2014LRSP...11....1P}. However, some essential aspects remain controversial. 

Prominence arcades consist of two parts. The spine is the horizontal and extended region parallel to the solar surface, and the two lateral connections between the spine and the underlying photosphere are the feet. Prominences are bright against the dark background of the sky, so the measured intensity profiles are in emission. We can also observe prominences on the disk (then called filaments). In that case, they appear dark against the bright Sun surface, with their intensity profiles in absorption. While prominences and filaments refer to the same type of structure, we focus on the case in which these structures appear as the former.

Prominences overarch the polarity inversion line of photospheric magnetic fields. Depending on the surrounding region, they are usually classified as quiescent or active. While the former appear in quiet Sun regions and slowly evolve over days, the latter occur near active regions (ARs) and evolve much faster \citepads{1974GAM....12.....T, 1995ASSL..199.....T, 2015ASSL..415.....V}. Regardless of their type, prominences are suspended at coronal heights, at which the plasma within the prominence is denser and cooler than the surroundings. 

Magnetic fields in the corona support the plasma in prominences against gravity. Observational studies usually reveal magnetic fields of 25--70~G in prominences \citepads[e.g.,][]{2001A&A...375L..39P, 2003ApJ...582L..51L, 2014A&A...566A..46O}. Although differences exist in estimating the field strength, it is the detailed characterization of the tridimensional magnetic geometry in prominences that is a much-disputed topic. While most observational studies concluded that prominences host rather horizontal fields \citepads[e.g.,][]{1983SoPh...83..135L, 1984A&A...131...33L, 2002Natur.415..403T, 2003ApJ...598L..67C, 2014A&A...566A..46O, 2019A&A...629A.138K, 2020A&A...644A..89D}, \citetads{2006ApJ...642..554M} presented some evidence for nearly vertical fields in a polar crown prominence\footnote{These authors clarified that if the aspect angle, $\delta$ (see their Fig. 3), during their observation was larger than $15^{\circ}$ it would be impossible to distinguish between nearly vertical and nearly horizontal magnetic fields, but from synoptic H-$\mathrm{\alpha}$ maps they presented evidence that the observed polar crown prominence was moving parallel to the southern limb during the observing period (i.e., that ${\delta}\, {<}\, 15^{\circ}$).}. This discrepancy is also patent in theoretical studies attempting to determine the vector magnetic field in prominences. Some models predict vertical fields in prominence spines \citepads[e.g.,][]{1989ApJ...336.1041O, 2000SoPh..194...73P}. On the other hand, others identify prominences as flux ropes where the spine is oriented parallel to the surface below and sustained by magnetic dips due to diverse causes, such as plasma pileup \citepads{1957ZA.....43...36K}, or the sheared configuration of the arcade \citepads{1994ApJ...420L..41A}, among other scenarios. Although none of the theoretical predictions can explain all of the observational constraints so far, this last set of models proposing that magnetic dips sustain the material of prominences is the one most supported by observations.

Furthermore, the solar limb or other structures appearing there can hide the prominence feet, severely affecting the determination of their magnetic field vector. An alternative to overcome this issue is using photospheric magnetic field measurements. \citetads{2006A&A...456..725L} employed this approach and reported the presence of local horizontal dips sustaining the prominence spine. However, \citetads{1998Natur.396..440Z} related prominence feet to locations vertically linking the magnetic field of the prominence spine to the photosphere. Specifically, this latter magnetic geometry may be compatible with the vertical and twisted magnetic ropes observed in the feet of a particular case of prominences, the so-called solar tornadoes \citepads{1932ApJ....76....9P}. Nonetheless, studies about these specific structures have led to contradictory results. \citetads{2015IAUS..305..275S} essentially find horizontal magnetic fields in a solar tornado and other prominences. Relatedly, \citetads{2023SSRv..219...33G} point out that solar tornadoes and other prominences have a similar magnetic geometry, so the differences previously reported in solar tornadoes might be due to projection effects. In contrast, \citetads{2015ApJ...802....3M} report helical magnetic fields perpendicular to the solar limb that link the prominence to the underlying atmosphere.

Observations in the \ion{He}{i} multiplets at 5876 (D$\mathrm{_{3}}$) and 10830~\AA\, are typically exploited to investigate the magnetic field in prominences. This is because the combined action of the Hanle and Zeeman effects makes the \ion{He}{i} multiplets a powerful diagnostic tool for characterizing the dynamic and magnetic properties of the plasma embedded in the chromosphere and corona \citepads[e.g.,][]{2022ARA&A..60..415T}. Specifically, the \ion{He}{i} D$\mathrm{_{3}}$ multiplet originates from the transition between the energy levels 1s2p $\mathrm{ ^{3}P_{0,1,2}}$ and 1s3d$\mathrm{^{3}D_{1,2,3}}$ of neutral helium. According to theoretical studies \citepads[e.g.,][]{1975ApJ...199L..63Z, 2008ApJ...677..742C}, this \ion{He}{i} multiplet system is populated through the photoionization-recombination mechanism. In this process, neutral helium in the chromosphere is ionized by incoming extreme ultraviolet (EUV) irradiation from the corona. Then, the ionized \ion{He}{ii} recombines with free electrons populating the multiplet system. For more information related to the \ion{He}{i} spectral lines, we refer the reader to, for example, \citetads{2002Natur.415..403T}, \citetads{2007ApJ...655..642T}, \citetads{2008ApJ...677..742C}, and references therein.

\begin{figure}[!t]
\centering
\includegraphics[width=0.5\textwidth, trim={0cm 0.2cm 2cm 2cm}, clip]{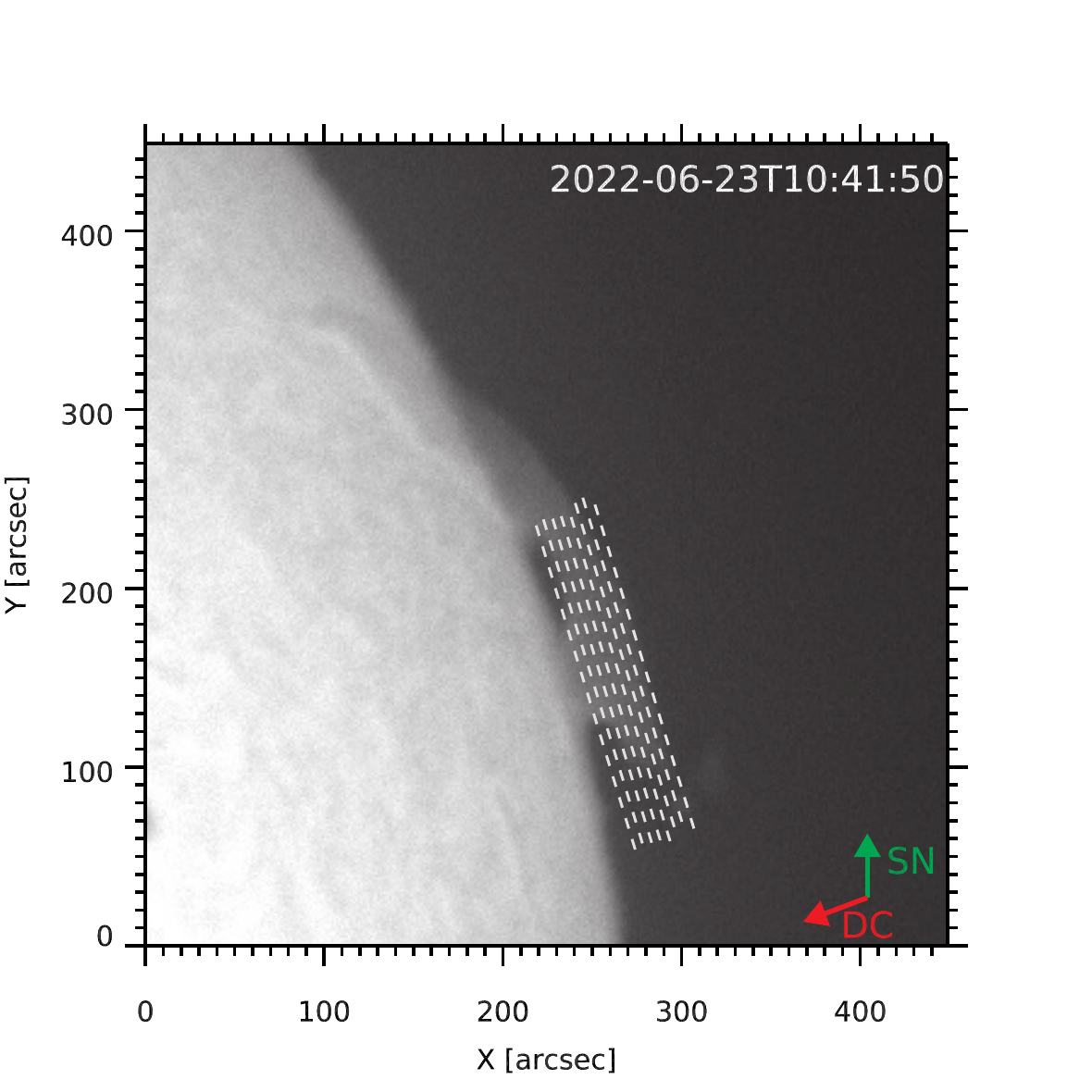}
\caption{Slit-jaw image from the secondary H-$\mathrm{\alpha}$ telescope at IRSOL showing the positions of the slit during the observation. The time stamp in the upper right corner shows the acquisition time of the H-$\mathrm{\alpha}$ image. The arrows in the lower right corner point to the disk center (DC) and solar north (SN).}
\label{fig:data}
\end{figure}

\begin{figure*}[!t]
\centering
\includegraphics[width=1.\textwidth, trim={2cm 0cm 1.5cm 2cm}, clip]{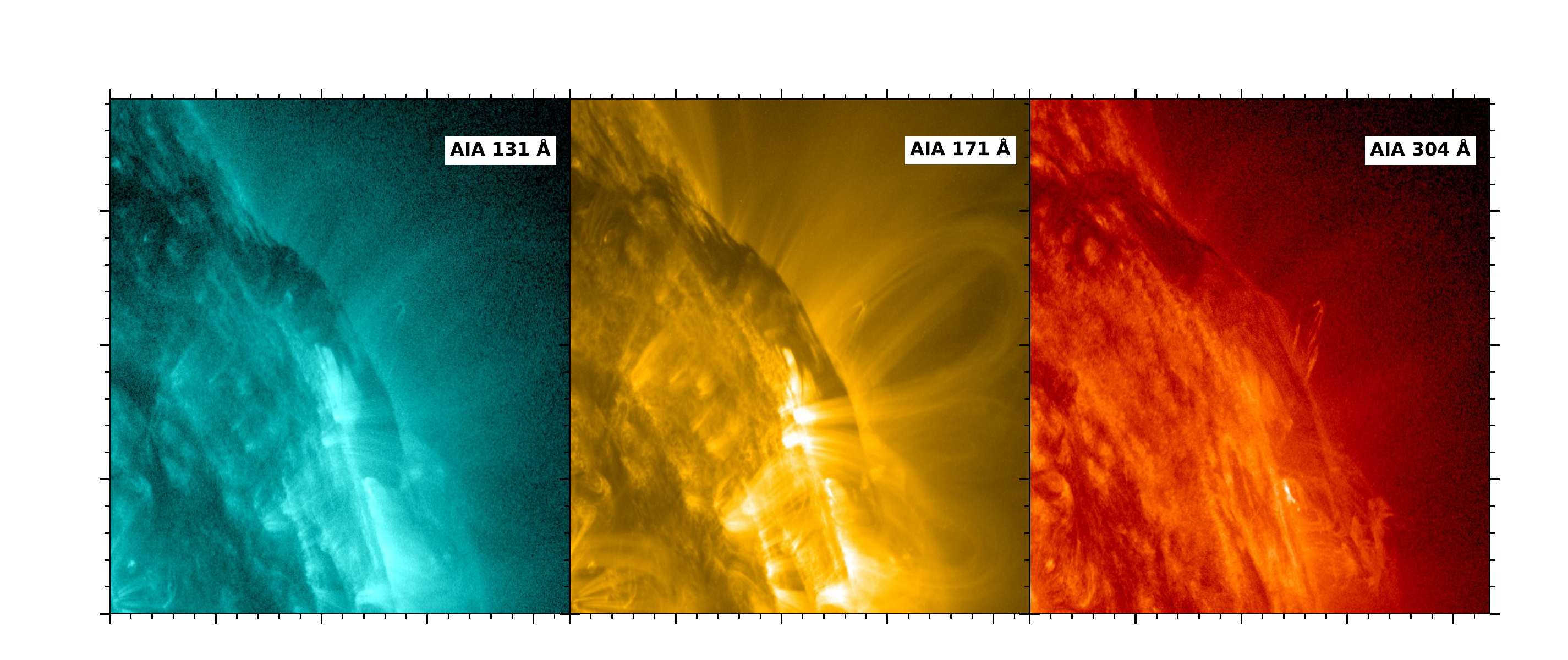}
\caption{Close-ups of SDO/AIA images showing the analyzed prominence. From left to right: SDO/AIA images in the 131~\AA, 171~\AA, and 304~\AA \, channels. These images were obtained at $\sim$11:55 UTC. Each major tick mark represents 100\arcsec.}
\label{fig:contexto}
\end{figure*}

In summary, prominences provide a prime opportunity to comprehend the magnetic field in the corona and its interaction with plasma. Analyzing prominences, however, poses a challenge due to the physical conditions in the surrounding corona and the intricate effects that need to be kept in mind during their study. Despite these liabilities, significant progress has been achieved over the years, though some key aspects remain unclear, and more studies are needed. On top of that, AR prominences have barely been investigated, so we know little about their properties compared to quiescent ones. 

Therefore, our investigation aims to shed light on the physical characterization of an AR prominence observed in the \ion{He}{i}~D$\mathrm{_{3}}$ multiplet at high polarimetric sensitivity. In the following, we outline the content of this paper. Section~\ref{sec:observations} describes the acquisition of the analyzed spectropolarimetric observations with the Z\"urich Imaging Polarimeter-3 \citep[ZIMPOL-3;][]{2010SPIE.7735E..1YR} and the data reduction. Then, in Sect.~\ref{sec:spectra}, we report interesting features detected in the Stokes profiles within the prominence. Section~\ref{sec:analysis} details the inversion strategy followed in this work, while we analyze the obtained results in Sect.~\ref{sec:results}. Finally, Sect.~\ref{sec:conclusions} outlines our results and presents relevant aspects to be considered.
  
\section{Observations and data reduction}
\label{sec:observations}

On June 23, 2022, between 08:34 and 12:33 UTC, we acquired full-Stokes measurements of an AR prominence in the \ion{He}{i}~D$_{3}$ multiplet using ZIMPOL-3, which is attached to the 45-cm aperture IRSOL Gregory Coud\'{e} Telescope \citep{2009ASPC..405...17B}, and a Czerny-Turner spectrograph. The observational conditions were good during the data acquisition, although influenced by high-altitude seeing. 

We used the ZIMPOL-3 system to perform this observation because it provides spectropolarimetric measurements at high polarimetric precision, which can be better than 10$\mathrm{^{-4}}$ with long exposure times. This advantage is given by the fast modulation rate of its photoelastic modulator \citep[PEM;][]{1997A&A...328..381G} and the synchronous on-chip demodulation of ZIMPOL-3. In combination with the fast modulation of the PEM, we also used a technique based on slow modulation to improve the zero-level accuracy of our polarimetric measurements, as it helps suppress the systematic instrumental polarization signals, in other words minimize $V$ $\rightarrow$ $Q$, $U$ crosstalk, which is relevant during that time of the year. This slow modulation is produced by a zero-order retarder mounted before the telescope aperture on the Telescope Calibration Unit (TCU). We refer the reader to \citet{2022SPIE12184E..0TZ} for more details about the slow modulation technique. With this setup, we acquired full-Stokes spectra.

The target structure was linked to AR 13032 during its course across the solar disk. At the time of the observation, a significant fraction of the structure was located on the west solar limb and another part was still visible against the disk, as is shown in Fig.~\ref{fig:data}. For context information, Fig.~\ref{fig:contexto} displays EUV images obtained with the Atmospheric Imaging Assembly \citepads[AIA;][]{2012SoPh..275...17L} of the Solar Dynamics Observatory \citepads[SDO;][]{2012SoPh..275....3P}, revealing the proximity of the magnetic loops of the AR close to the prominence during the observation.

We obtained spectropolarimetric data from most of the spine and part of a foot of the prominence. We performed sit-and-stare sequences at different limb distances. The spectrograph slit was oriented parallel to the limb and moved by 5\arcsec\,perpendicular to it. Specifically, Fig.~\ref{fig:data} shows the position of the slit in each sequence (dashed white lines). Considering the performance of the setup used, each sequence comprises on the order of 3100 frames, where intensity, circular, and linear polarization measurements are in equal number \citepads[see Sect. 4 of ][]{2022SPIE12184E..0TZ}. The spectral images cover a range of 10.45~\AA \, with a spectral sampling of 8.43~m\AA~pixel$\mathrm{^{-1}}$. The integration time was 2~s~frame$\mathrm{^{-1}}$ for measurements at 5--25\arcsec \,from the limb and 3~s~frame$\mathrm{^{-1}}$ for those performed further away. The orientation of the scans was from solar south to solar north along the slits displayed in Fig.~\ref{fig:data}. The slit width was 80~$\mu$m ($\sim$0.65\arcsec) and the pixel scale of the ZIMPOL-3 camera $\sim$1.3\arcsec~pixel$\mathrm{^{-1}}$.

The TCU compensates for the $I$ $\rightarrow$ $Q, U, V$ and $V$ $\rightarrow$ $Q, U$ crosstalks. To determine how the $Q, U$  $\rightarrow$ $V$ crosstalk affects our data, we need to consider that these terms influence our data by a factor (1 + cos($\delta$)), where $\delta$ is the retardance of the TCU \citepads{2022SPIE12184E..0TZ}. In our case, this factor is $\sim$0.05. The $Q$ $\rightarrow$ $U$ crosstalk by the telescope cannot be corrected with the TCU. According to the theoretical modeling of the Mueller matrix of the telescope, the crosstalk $Q$ $\rightarrow$ $U$ (and viceversa) generated by the telescope is not expected to exceed 1\%.

Calibration data were taken regularly during the observing run. Specifically, we acquired flat field data after each sit-and-stare sequence. We also performed background measurements outside the prominence (at $\sim$100\arcsec\,from the limb) to compensate for the intensity and polarization in our observations due to straylight contamination. This particular contamination is due to spurious illumination of the camera, which can severely affect off-limb observations.

Data reduction was performed following the standard process for ZIMPOL-3 data. First, we corrected for dark images. Then, we calibrated polarimetrically and corrected for flatfield. Concretely, we used the calibration measurements obtained closest in time to each observed sequence. In addition, we applied a curvature correction to the Q/U coordinate system so that the resulting +$Q$ direction is parallel to the closest limb. Finally, we compensated for straylight contamination by subtracting the background measurements from our observations, as was described in \citetads{2012SoPh..281..697R}.

In order to increase the signal-to-noise ratio of our observations, the data were binned spatially and spectrally by 4. After binning, the average noise levels for $I\mathrm{/}I_{\mathrm{max}}$, $Q\mathrm{/}I_{\mathrm{max}}$, $U\mathrm{/}I_{\mathrm{max}}$, and $V\mathrm{/}I_{\mathrm{max}}$ are 3.35$\times$10$^{-3}$, 9.73$\times$10$^{-5}$, 9.72$\times$10$^{-5}$, and 8.42$\times$10$^{-5}$, respectively, measured in the continuum close to the \ion{He}{i} D$_{3}$ line. $I_{\mathrm{max}}$ represents the maximum intensity signal along each slit. Hereafter, the data shown in this paper correspond to the binned spectra and, for readability, we shall omit the division by $I_{\mathrm{max}}$ when addressing the Stokes parameters throughout the text.

\begin{figure*}[!t]
\centering
\includegraphics[width=1.\textwidth, trim={0cm 0cm 2cm 1cm}, clip]{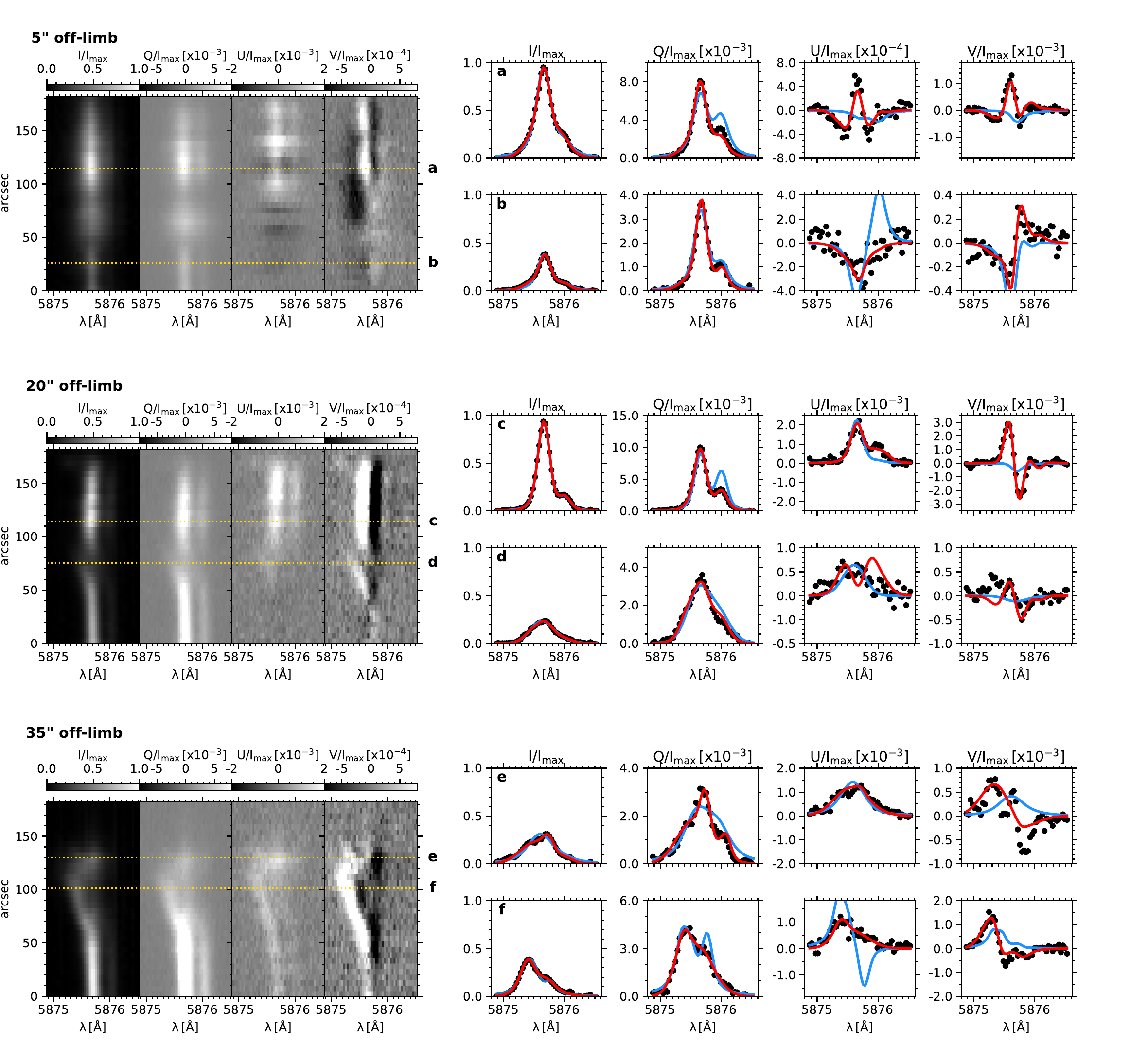}
\caption{Details of the spectropolarimetric data and synthetic profiles inferred from the inversions. Left: Spectra emerging along the slits located at 5\arcsec, 20\arcsec, and 35\arcsec from the solar limb. Right: Stokes profiles at the positions labeled a--f and marked with dashed yellow lines on the left panels (dotted lines). The synthetic profiles obtained when assuming one and two magnetic components are represented by blue and red lines, respectively. I$\mathrm{_{max}}$ stands for the maximum intensity signal along each slit.}
\label{fig:espectros_perfiles_ajustes}
\end{figure*}

\section{Characteristics of the Stokes spectra}
\label{sec:spectra}

Our datasets contain an impressive variety of Stokes profiles, giving us a first impression of the vast amount of information unveiled in the observed prominence. Specifically, Fig.~\ref{fig:espectros_perfiles_ajustes} outlines the appearance of the observed Stokes profiles emerging along three slit positions at different limb distances. We remind the reader that the bottom (top) part of the spectra was measured at the south (north) end of each slit shown in Fig.~\ref{fig:data}.

At 5\arcsec\,off-limb (top panels of Fig.~\ref{fig:espectros_perfiles_ajustes}), the Stokes profiles are practically unshifted, and the blue and red components of the spectral line are discernible, overall in Stokes $Q$, as in \citetads{2005ASSL..320..215R}. The measured spectra along the slit show conspicuous changes. In general, the amplitudes of all Stokes profiles increase along the slit (that is, as we move toward the prominence). Stokes $Q$ and $U$, moreover, show different shapes varying along the slit, probably due to changes in the fine structure of the prominence. Stokes $V$ also displays a striking variation, where the weak two-lobed Stokes $V$ profiles detected outside the prominence turn progressively into signals with one predominant blue lobe, three lobes, and two lobes within the structure. This variation in the circular polarization inside the prominence may indicate a change from weak to stronger fields; that is, from a field regime with a dominant alignment-to-orientation mechanism \citepads{1984ApJ...278..863K} to another one that is under the predominant influence of the Zeeman effect.

At 20\arcsec\,from the solar limb (middle panels of Fig.~\ref{fig:espectros_perfiles_ajustes}), the Stokes $I$ and $Q$ profiles are practically homogeneous along the slit, except for an abrupt blueshift at 60--90\arcsec where they are broader and weaker (see panels labeled d). On the other hand, Stokes $U$ is weak at the beginning of the slit and increases significantly between 110 and 165\arcsec. Specifically, this increase in Stokes $U$ comes together with a decrease in $Q$, suggesting a change in the magnetic configuration of the prominence at such positions. Furthermore, the Stokes $V$ signals show different changes along the slit. We first observe weak two-lobed Stokes $V$ profiles whose polarity changes at about 40\arcsec. At 60--90\arcsec, these profiles become more complex as only the blue lobe is shifted. After that, the amplitude of the lobes increases significantly, indicating a dominant influence of the Zeeman effect.

The bottom panels of Fig.~\ref{fig:espectros_perfiles_ajustes} show conspicuous lineshifts at 35\arcsec \,away from the solar limb. There, all Stokes profiles are strongly blueshifted between 70 and 110\arcsec. This fact suggests strong flow motions inside the prominence whose Doppler velocities can reach up to $-$15~km~s$\mathrm{^{-1}}$. At such positions, the emerging Stokes profiles are shifted and wide (see profiles f). In particular, Stokes $V$ has a complex shape formed by a shifted asymmetric profile and an extra lobe in the red wing. We can distinguish other significant changes along the slit. Before the lineshift, the $I$ and $Q$ profiles weaken, while $U$ and $V$ increase. The variation in the profile amplitudes along the slit may imply changes in the configuration and intensity of the magnetic field in the prominence: the former is suggested by the variation in the $Q$ and $U$ signals and the latter by the increase in Stokes $V$. After the lineshift, Stokes profiles show intricate shapes (see profiles e) until they vanish outside the prominence.

\section{Analysis}
\label{sec:analysis}

We inferred the thermodynamic and magnetic properties of the observed prominence by inverting our spectropolarimetric observations with the Hanle and Zeeman light inversion code\footnote{Publicly available on https://aasensio.github.io/hazel2} \citep[HAZEL;][]{2008ApJ...683..542A}. In the inversion process, we assumed that the prominence was in the plane of the sky during the observing period. 

The HAZEL code is able to perform synthesis and inversions of Stokes profiles produced by the joint action of the radiatively induced atomic level polarization and the magnetic field through the Hanle and Zeeman effects. Specifically, HAZEL assumes a cloud of \ion{He}{i} atoms (hereafter, slab) placed at a certain height above the solar surface. This slab has an optical thickness, $\tau$, and constant physical properties. In particular, HAZEL does not explicitly take into account the radiative mechanism that is assumed to be responsible for the overpopulation of the multiplet levels of \ion{He}{i} required to produce the absorption or emission features observed in the spectral lines of the 10830 and D$\mathrm{_{3}}$ multiplets. Instead, the overpopulation is taken care of by the optical depth parameter needed to fit the Stokes $I$ profiles. 

Moreover, the code assumes that these atoms in the multiplet levels are illuminated from below by the photospheric continuum radiation field, and that this is the only anisotropic incident radiation that produces atomic level polarization in the \ion{He}{i} levels at the height corresponding to the spatial point being considered. This radiation field is computed from the tabulated center-to-limb variation in the continuum intensity \citep{Pierce2000}. Once the population and the atomic polarization of the \ion{He}{i} levels is determined through the solution of the statistical equilibrium equations, the emergent Stokes profiles are obtained by solving the Stokes-vector transfer equation, which in a constant-property slab has an analytical solution \citepads{2005ApJ...619L.191T}. For more details about HAZEL, we refer the reader to \citetads{2008ApJ...683..542A}.

\begin{table}
\caption{Initialization of the inversions performed along each slit.}             
\label{table:initialization_inversions}      
\centering                          
\begin{tabular}{l c c c c c c}        
\hline\hline                 
Inv. (com.) & ($\mathrm{B_{x}, B_{y}, B_{z}}$) & $\tau$ & $\mathrm{v_{LOS}}$ & $\mathrm{\Delta v}$ \\    
  & (G) & & (km~s$\mathrm{^{-1}}$) & (km~s$\mathrm{^{-1}}$) \\
\hline                        
  1c (1) & (0.2, 5.0, $-$0.1) & 0.1 & 2.0 & 4.0 \\   
\hline
  2c (1) & (0.2, 5.0, $-$0.1) & 0.1 & 2.0 & 4.0 \\  
  2c (2) & (0.2, 2.0, $-$0.1) & 0.1 & 2.0 & 10.0 \\ 
\hline                                   
\end{tabular}
\tablefoot{Each column represents the inversion mode (component), magnetic field vector components, optical depth, LOS velocity, and thermal broadening.}
\end{table}

\begin{figure}[!t]
\centering
\includegraphics[width=0.42\textwidth, trim={0.5cm 0.3cm 1.2cm 0.9cm}, clip]{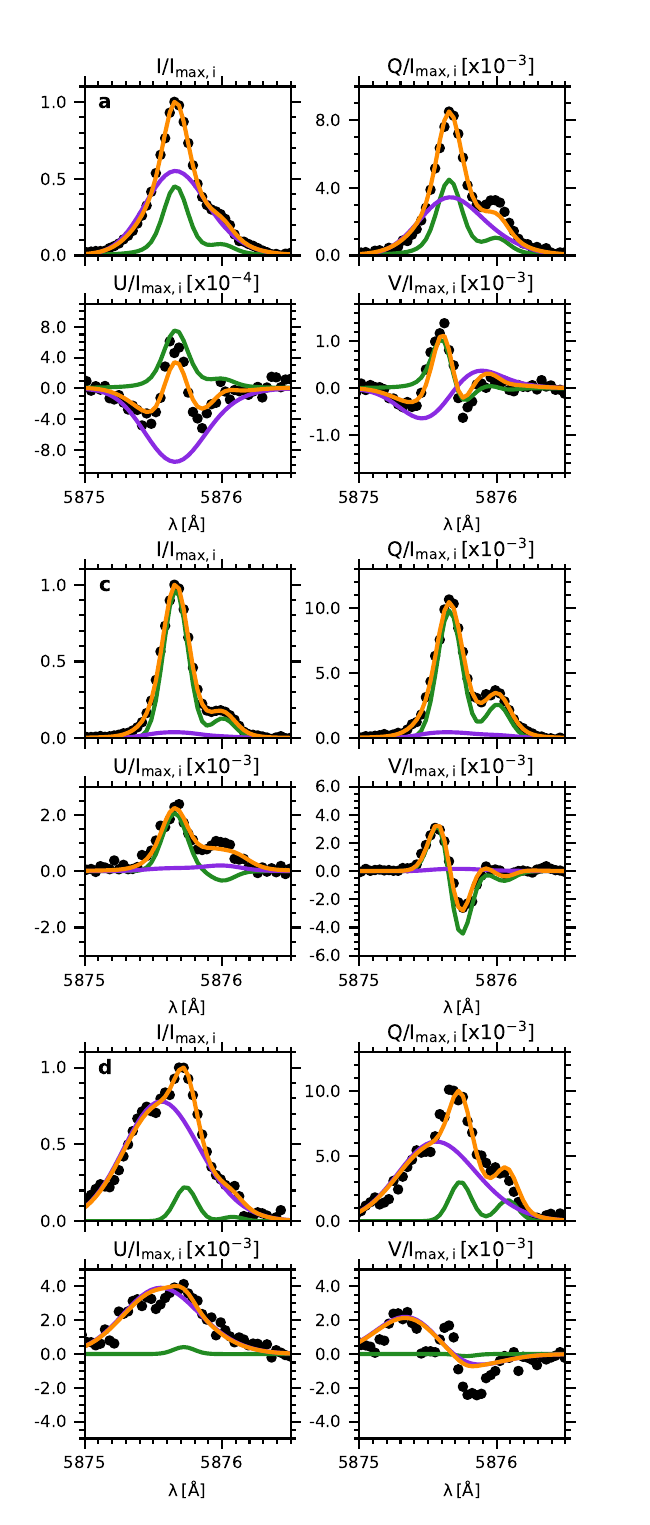}
\caption{Synthetic profiles resulting from the two atmospheric components (green and violet, respectively) assumed in the inversion of the Stokes profiles labeled a, c, and d in Fig.~\ref{fig:espectros_perfiles_ajustes}. The observed and best-fit profiles are shown, respectively, in black (dotted curves) and orange. I$\mathrm{_{max, i}}$ stands for the maximum intensity signal of the plotted pixel.}
\label{fig:synthesis}
\end{figure}

The data acquired during each slit sequence was inverted independently by assuming that the slab of \ion{He}{i} atoms was placed in the plane of the sky at a height equal to the distance of each slit to the solar limb. We followed a two-step strategy to perform the inversions of the binned Stokes profiles using a one-component model. During the first cycle, we only considered the Stokes~$I$ profiles to retrieve the thermodynamical parameters, which are the optical depth at the wavelength 5876~\AA\,($\tau$), line-of-sight (LOS) velocity ($\mathrm{v_{LOS}}$), thermal broadening ($\mathrm{\Delta v}$), and damping parameter (a). In the second cycle, we only used the polarization profiles to infer the magnetic field vector components ($\mathrm{B_{x}}$, $\mathrm{B_{y}}$, and $\mathrm{B_{z}}$). Each slit was inverted in serial mode; that is, we used the output atmospheric model from one pixel as the input model of the next one. Since the first pixel was located outside the prominence (see coordinates (X, Y) $\sim$ (290\arcsec, 70\arcsec) in Fig.~\ref{fig:data}), we initialized the inversion of each slit by using an input model with a low magnetic field, LOS velocity, and thermal broadening values (listed on the first row of Table \ref{table:initialization_inversions}). We proved the suitability of this initial model by comparing observed and synthetic profiles. The used inversion strategy allowed us to speed up convergence and to keep consistency among the results obtained along each slit. 

The blue curves in the right-hand panels of Fig.~\ref{fig:espectros_perfiles_ajustes} show the synthetic profiles resulting from the one-component inversions. Only intensity profiles display a good fit quality, indicating that a one-component model is insufficient to explain most of the polarization profiles. 

To improve the fit quality of the polarization profiles, we repeated our inversion strategy considering two magnetic components, which were assumed to lie side by side inside the pixel. Table~\ref{table:initialization_inversions} lists the corresponding initial input models. The filling factor used to initialize the two-component inversion was 50\% in both components. We also proved the suitability of these initial models as we did for the one-component model. The resulting synthetic profiles usually match both intensity and polarization profiles (red curves in Fig.~\ref{fig:espectros_perfiles_ajustes}). Thus, assuming the coexistence of two magnetic components appears to be an adequate approach to explain our observations as the synthetic profiles can reproduce many of the details found in the observed profiles, such as their width, shape, and the number of lobes in Stokes $V$. The quality of the fit of some polarization profiles is still low because more complex atmospheric models are probably needed.

When considering a two-component model, each atmosphere contributes differently to the output synthetic profiles that HAZEL provides. As an example, Fig.~\ref{fig:synthesis} portrays the contribution of each component to three of the synthetic profiles shown in Fig.~\ref{fig:espectros_perfiles_ajustes}. Hereafter, these components are labeled 1 and 2 (plotted, respectively, in green and violet). Despite the different cases presented in Fig.~\ref{fig:synthesis}, component 1 usually contributes more significantly to the output synthetic profiles than component 2 (see Sect.~\ref{sec:results}). However, combining both components is essential to give shape to the output synthetic profiles (orange curves). The synthetic profiles of component 2 are wider as a common rule, and are responsible for broadening the output synthetic profiles. Meanwhile, component 1 contributes to other line features in the output synthetic profiles. Combining both components, moreover, reproduces three-lobed Stokes $V$ profiles. 

Although the fit quality between the observed and synthetic profiles improves when using a two-component model, we need to discern the output model that best suits each position. To do so, we computed the Bayesian information criterion \citep[BIC;][]{1978AnSta...6..461S}, which has been proven successfully in other studies in solar physics (e.g., \citealt{2017A&A...608A..97F}, \citealt{2023A&A...671A..79C}). The BIC represents a valuable parameter for comparing models when the number of observed points ($N$) exceeds the number of free parameters ($k$), and is computed as

\begin{equation}
\mathrm{BIC} = \chi^2_{\mathrm{min}} + k \ln N,
\label{eq:bic}
\end{equation}

\noindent where $N$ is the number of wavelength positions multiplied by four, since we are using full-Stokes measurements, and $\chi^2_{\mathrm{min}}$ is the merit function:

\begin{equation}
\chi^2_{\mathrm{min}} = \sum_{j=1}^{N} \left(\dfrac{s_{j}-o_{j}}{\sigma_{j}}\right)^2,
\end{equation}

\noindent $s_{j}$ being the best-fit profiles given by HAZEL, $o_{j}$ the observed profiles, and $\sigma_{j}$ the noise level estimated in the continuum. The advantage of using the BIC parameter is that it not only compares the $\chi^2_{min}$ value given by each model but also balances this comparison, penalizing the models with more free parameters. In our study, the one- and two-component inversions had, respectively, 7 and 16 free parameters. 

\begin{figure}[!t]
\centering
\includegraphics[width=0.495\textwidth, trim={0.2cm 0.1cm 0.3cm 0.1cm}, clip]{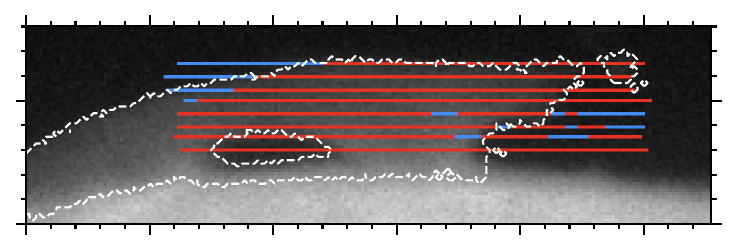}
\caption{Model selection along each slit overplotted on a rotated H-$\alpha$ slit-jaw image of the analyzed prominence. Locations described by one- and two-component models are displayed, respectively, in blue and red. The dashed white line encloses the prominence. Each major tick mark represents 50\arcsec.}
\label{fig:conv_modelo}
\end{figure}

After computing the BIC parameter for each inverted position, we selected the model with the lowest BIC as the most appropriate one at each position. Figure~\ref{fig:conv_modelo} shows the model selection along each slit. For display purposes, the slit is broader and does not represent the actual slit width during the observations. We found that the two-component model describes 83.2\% of the positions. The one-component model is usually related to positions outside the prominence or with noisy polarization profiles. 

\section{Results}
\label{sec:results}

\begin{figure*}[!t]
\centering
\includegraphics[height=0.83\textheight, trim={0cm 0cm 0cm 0cm}, clip]{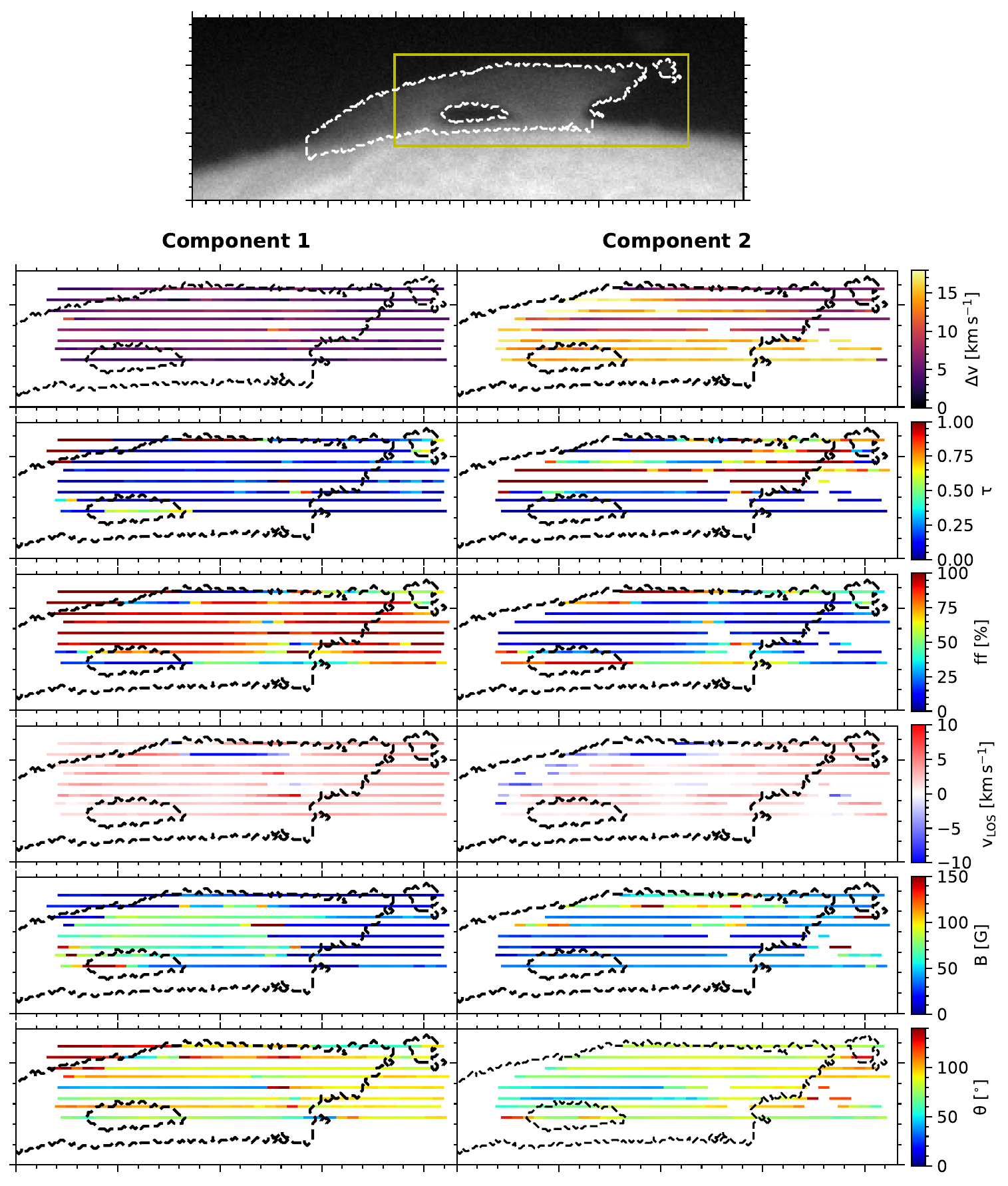}
\caption{Inversion results given by the HAZEL code along each slit. The upper panel shows a rotated H-$\mathrm{\alpha}$ slit-jaw image of the analyzed prominence, which is enclosed by a dashed white line. The lower panels show the inferred physical quantities for both components at each slit position in the field of view delimited by a yellow rectangle in the upper panel. From top to bottom: Thermal broadening ($\mathrm{\Delta}$v), optical thickness ($\mathrm{\tau}$), filling factor (ff), LOS velocity (v$\mathrm{_{LOS}}$), magnetic field strength (B), and inclination ($\theta$). Each major tick mark represents 50\arcsec.}
\label{fig:results_inv}
\end{figure*}

After describing our inversions with HAZEL, we analyze the results retrieved along each slit in this section. Figure~\ref{fig:results_inv} displays the inferred values for the thermal broadening, optical thickness, filling factor, LOS velocity, magnetic field strength, and inclination of the magnetic field with respect to the vertical for components 1 and 2 over the observed area of the prominence. As in Fig.~\ref{fig:conv_modelo}, the slits shown are broader than the ones used in the observing run for visualization purposes. Blank locations in the component 2 panels represent positions described by only one component.

\begin{table*}[!t]
\label{table:median_iqr}
 \caption{Statistics on the physical quantities inferred for each component at different limb distances.}
 \centering
\begin{tabular}{cccccccccccccccc}
 \hline \hline
Limb & Com. & \multicolumn{2}{c}{$\Delta$v} & \multicolumn{2}{c}{$\tau$} & \multicolumn{2}{c}{ff} & \multicolumn{2}{c}{v$\mathrm{_{LOS}}$} & \multicolumn{2}{c}{B} & \multicolumn{2}{c}{$\mathrm{\theta}$}\\ 
distance & & med. & iqr & med. & iqr & med. & iqr & med. & iqr & med. & iqr & med. & iqr \\
(arcsec) & & (km s$\mathrm{^{-1}}$) & (km s$\mathrm{^{-1}}$) & & & (\%) & (\%) & (km s$\mathrm{^{-1}}$) & (km s$\mathrm{^{-1}}$) & (G) & (G) & ($^{\circ}$) & ($^{\circ}$) \\
\hline
5--10 & 1 & 4.44 & 1.68 & 0.01 & 0.01 & 0.67 & 0.47 & 2.54 & 1.10 & 36.61 & 41.72 & 88.76 & 23.79\\
 & 2 & 14.97 & 1.60 & 0.02 & 0.02 & 0.33 & 0.47 & 1.36 & 1.32 & 38.17 & 5.71 & 78.66 & 11.54\\
15--20 & 1 & 6.13 & 1.17 & 0.08 & 0.01 & 0.95 & 0.09 & 2.74 & 1.52 & 57.28 & 54.20 & 80.49 & 49.02\\
 & 2 & 10.75 & 8.11 & 0.25 & 1.31 & 0.05 & 0.09 & 1.21 & 2.32 & 16.05 & 20.26 & 54.86 & 43.66\\
25--30 & 1 & 5.84 & 1.73 & 0.09 & 0.06 & 0.92 & 0.07 & 3.14 & 0.87 & 66.29 & 57.06 & 87.53 & 12.05\\
 & 2 & 8.60 & 6.31 & 0.80 & 0.70 & 0.08 & 0.07 & 2.39 & 2.43 & 40.45 & 11.22 & 88.23 & 18.18\\
35--40 & 1 & 3.87 & 2.05 & 0.08 & 0.49 & 0.85 & 0.53 & 2.86 & 2.60 & 14.89 & 14.86 & 97.75 & 35.72\\
 & 2 & 6.56 & 7.15 & 0.50 & 0.87 & 0.15 & 0.53 & 0.05 & 5.80 & 42.12 & 71.80 & 81.91 & 4.60\\
\hline
\hline
\end{tabular}
\tablefoot{Median (med.) and interquartile ranges (iqr) of the thermal broadening ($\Delta$v), optical thickness ($\tau$), filling factor (ff), LOS velocity (v$\mathrm{_{LOS}}$), magnetic field strength (B), and inclination of the magnetic field vector ($\theta$) inferred for components 1 and 2 depending on the limb distance.}
\end{table*}

\begin{figure}[!t]
\centering
\includegraphics[width=0.48\textwidth, trim={0.7cm 1cm 2.2cm 1.9cm}, clip]{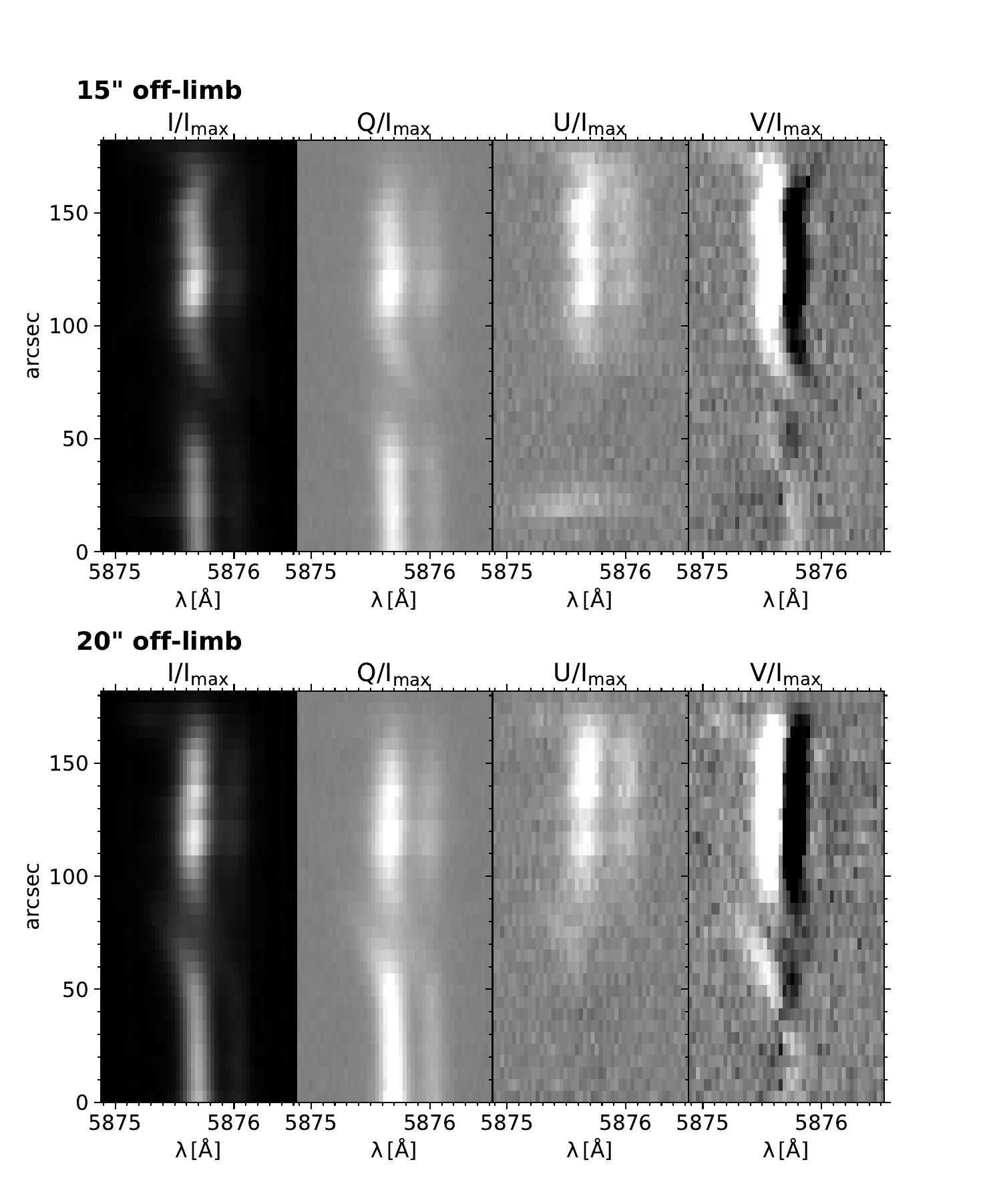}
\caption{Stokes profiles along the slits located at 15\arcsec \,and 20\arcsec \,from the solar limb. The ranges used for color saturation in the I/I$\mathrm{_{max}}$, Q/I$\mathrm{_{max}}$, U/I$\mathrm{_{max}}$, and V/I$\mathrm{_{max}}$ panels are [0, 1], [$-$8$\times$10$\mathrm{^{-3}}$, 8$\times$10$\mathrm{^{-3}}$], [$-$2$\times$10$\mathrm{^{-3}}$, 2$\times$10$\mathrm{^{-3}}$], and [$-$8$\times$10$\mathrm{^{-4}}$, 8$\times$10$\mathrm{^{-4}}$], respectively. I$\mathrm{_{max}}$ stands for the maximum intensity signal along each slit.}
\label{fig:shifts}
\end{figure}

First, we describe the results shown in Fig.~\ref{fig:results_inv} for each component. Table~2 enumerates statistics on the physical quantities inferred at different limb distances. In particular, we opted to compute the median and interquartile ranges because not all the listed parameters follow a Gaussian distribution.  

Prominences reveal emission profiles with a substantial broadening due to thermal and turbulent velocities. Our inversions usually provided broader synthetic profiles for component 2 so, accordingly, this component has a significantly larger thermal broadening compared to that of component 1. Specifically, the thermal broadening is 1--6~km~s$\mathrm{^{-1}}$ for component 1 and can exceed 7 km~s$\mathrm{^{-1}}$ for component 2 (see first row of Fig.~\ref{fig:results_inv}). Only component 2, moreover, shows a decrease in thermal broadening with the distance to the limb, while we do not detect a clear variation for component 1. In particular, the values obtained for component 1 at all limb distances and component 2 at 25--40\arcsec\,are similar to the ones previously reported in prominences \citepads[e.g.,][]{2011ASPC..437..109R, 2014A&A...566A..46O}. Meanwhile, close to the limb, component 2 shows a median of almost 15~km~s$\mathrm{^{-1}}$, which is compatible with the thermal broadening found in spicules by \citetads{2005ApJ...619L.191T}. The origin of such a high thermal broadening close to the limb may be related to strong dynamical processes in the plasma existing there \citepads[e.g.,][]{2007ASPC..368..291L}. 

We estimated the temperature for each component considering the inferred thermal broadening values. As was expected, there is a substantial difference between the temperature obtained for each component. The temperature of component 1 is usually 9--25$\times$10$\mathrm{^3}$~K, which is comparable to temperatures typically found in prominences \citepads[e.g.,][]{2014LRSP...11....1P}. Component 2 is much hotter (with temperatures on the order of 10$\mathrm{^5}$~K), which suggests that the high Doppler broadening inferred for this component may have a nonthermal contribution that is noteworthy. 

Another plausible explanation for the high temperatures inferred for component 2 could be the imprint of a prominence-corona transition region (PCTR). Therefore, the temperature difference between the assumed components may be compatible with that in a prominence model comprising a cool prominence body with an inner PCTR region and the hot outer region of the PCTR \citepads{1999A&A...349..974A}. In particular, \citetads{2004ApJ...617..614L} reported on the importance of the PCTR in the formation of \ion{He}{i} multiplets in prominences.

We also found differences in the optical thickness retrieved for each atmosphere (second row of Fig.~\ref{fig:results_inv}). Component 1 mainly shows $\tau$ values below 1, and is only greater at positions with broad Stokes $I$ profiles at 30--40\arcsec\,off-limb. On the other hand, component 2 has a wider variety of optical thickness values. In particular, this latter component is related to a thicker atmosphere compared to component 1 at distances further than 15\arcsec\,from the solar limb, where $\tau$ values are significantly above 1.

Regarding the filling factor, component 1 usually shows greater values (third row of Fig.~\ref{fig:results_inv}), with a median of 70--90\%. The fact that, on average, we retrieved lower filling factors for component 2 reinforces the idea that properties obtained for this component may be related to the presence of a PCTR. At the same time, we also found positions where the filling factor is either similar for both components or greater for component 2. In particular, this occurs more frequently close to the limb, where Stokes profiles are broader and component 2 has a significant contribution.

Previous studies revealed that prominences host dynamic flows \citepads[e.g.,][and references therein]{2010SSRv..151..243L, 2014LRSP...11....1P, 2015ASSL..415...79K, 2023PhDT.........1P}, which are related to LOS velocities of 5--25 km~s$\mathrm{^{-1}}$ in active prominences. Our results show more moderate LOS velocities that are analogous to those found in quiescent prominences \citepads[e.g., ][]{2010A&A...514A..68S, 2012ApJ...761L..25O, 2014A&A...566A..46O, 2015ApJ...802....3M}. Specifically, we inferred LOS velocities of 2--3~km~s$\mathrm{^{-1}}$ and $-$3--1~km~s$^{-1}$, respectively, for components 1 and 2. Moreover, component 2 shows blueshifted LOS velocities between $-$1 and $-$5~km~s$\mathrm{^{-1}}$ that are possibly related to brightenings disclosed in the prominence by SDO/AIA images (see Fig.~\ref{fig:contexto}).

Other interesting features show up in the LOS velocity maps. First, both components reveal locations where the LOS velocity reaches up to 16~km~s$\mathrm{^{-1}}$. These potentially supersonic\footnote{The typical sound speed at the chromosphere is 10~km~s$\mathrm{^{-1}}$.} LOS velocities appear at particular positions at 35--40\arcsec\,away from the solar limb where spectra are blueshifted (see the lower row in Fig.~\ref{fig:espectros_perfiles_ajustes}). By comparing with Fig.~\ref{fig:contexto}, these supersonic velocities may be related to flow motions in the near AR loops or even to small-scale flows in the prominence that eluded detection due to a lack of spatial resolution. 

Furthermore, both components show a transition from redshifted, blueshifted, to again redshifted LOS velocities at 15--25\arcsec\,off-limb (located at 125--150\arcsec\,in the horizontal axis of Fig.~\ref{fig:results_inv}). Figure~\ref{fig:shifts} displays the abrupt redshift and blueshift at the slit positions between 70\arcsec\,and 90\arcsec\,in the spectra measured, respectively, at 15 and 20\arcsec\,to the solar limb. Although the LOS velocities are redshifted again at 25\arcsec\,off-limb, the difference between the spectra at 20\arcsec\,and 25\arcsec\,to the solar limb is not as clear as that represented in Fig.~\ref{fig:shifts}, which is possibly due to fluctuations in the seeing conditions. In any case, this pattern of the LOS velocity reminds us of those found by \citetads{2014A&A...566A..46O} and \citetads{2015ApJ...802....3M}, who interpreted it as an indicator of the presence of twisted flows.

Regarding the magnetic field strength, we retrieved values of a few tens of gauss in both components, with a median of 15--65~G. These field strengths are compatible with those previously reported (see Sect.~\ref{sec:intro}). The median of the field strength of component 1 increases from 35 to 65~G with the distance to the limb, until it suddenly drops to $\sim$ 15~G at 40\arcsec\,to the limb. Meanwhile, component 2 shows field strengths with a median of $\sim$40~G, except at 15--20\arcsec\,off-limb where it weakens to about 15~G. We also found fields of 100--200~G standing out in both components. Commonly, these stronger fields appear at locations with strong and broad polarization signals. 

The retrieved magnetic field inclination indicates that the field inside the prominence is rather parallel to the solar surface (bottom row of Fig.~\ref{fig:results_inv}). That is, we inferred mainly horizontal fields. Specifically, we obtained inclinations of 70--100$\mathrm{^\circ}$ for both components, with a median of $\sim$90$\mathrm{^\circ}$ at all limb distances. In addition, we inferred more vertical magnetic fields at particular locations. For instance, both components show field inclinations of 60--70$\mathrm{^\circ}$ at 5\arcsec\,off-limb, suggesting more vertical fields close to the solar limb. Moreover, at 15--20\arcsec\,off-limb, the prominence spine shows field inclinations of $\sim$50$\mathrm{^\circ}$, suggesting the presence of vertical field lines pointing away from the solar surface. Component 1 also reveals field inclinations that can exceed 120$\mathrm{^\circ}$, indicating field lines that point to the solar surface at some particular positions, such as those showing potential supersonic velocities at 35\arcsec\,off-limb and a possible twist at 15--25\arcsec\,to the limb. Some positions at the edges of the slits also display inclination values greater than 120$\mathrm{^\circ}$ for both components.

Therefore, the prominence hosts vertical and horizontal magnetic fields, with the latter being predominant. These changes in the field orientation are spatially coherent. In other words, they appear as a conglomerate of homogeneous results along consecutive positions. At this point, we note that the information emerging from the prominence might be mixed with that from the near AR loops. In such a case, the differently oriented fields retrieved in the prominence might instead be related to the AR loops and the prominence. 

After describing the physical quantities obtained for each component, we analyze our results by considering them together. Component 1 is an optically thin atmosphere with smaller thermal broadening. On the other hand, component 2 has a significantly larger thermal broadening and is sometimes optically thicker. Indeed, the optical thickness inferred for component 2 sometimes exceeds 2--3, which is not expected when analyzing \ion{He}{i}~D$\mathrm{_{3}}$ data. At the same time, component 1 shows, in general,  the greatest filling factors.

Pursuing this last aspect further, component 2 shows greater $\tau$ values at locations where the filling factor is small. Component 2 can be considered a minor (but important) contributor to the output result given by HAZEL. Thus, the role of component 2 is to provide a broad enough synthetic profile that, after being combined with that from component 1, improves the fit with the observed profile.

Summarizing, we inferred a correlation of some physical quantities with the distance to the limb. The physical processes occurring in the prominence close to the limb can be explained by two atmospheric components that contribute almost equally to the output result and show similar physical properties. However, these components reveal significantly different thermal broadenings, reaching up to 15~km~s$\mathrm{^{-1}}$ in component 2. Such a large thermal broadening is expected as plasma located close to the limb can undergo strong dynamic processes. On the other hand, further from the solar limb, component 1 contributes predominantly to the output result. The physical processes occurring in the prominence can thus be explained by the parameter values inferred for this component, which reveal redshifted LOS velocities of 1--3 km~s$\mathrm{^{-1}}$ and mostly horizontal magnetic fields of 20--80~G. 

Furthermore, we found that the central region of the prominence spine harbors a magnetic field that is 20~G stronger than in other locations of the prominence. This variation is different to the finding of \citetads{2014A&A...566A..46O} in a quiescent prominence. These authors, moreover, reported a correlation between field strength and optical depth at the wavelength 10830~\AA. In contrast, we have not found a clear change in the optical depth (considering the values for component 1) with the magnetic field strength.

\section{Conclusions}
\label{sec:conclusions}

Analyzing solar prominences is an engaging topic that broadens our knowledge of the properties and the behavior of plasma and magnetic fields in the corona. Our study brings more information and ideas to our understanding of prominences. We used spectropolarimetric data of an AR prominence in the \ion{He}{i}~D$\mathrm{_{3}}$ multiplet obtained with the ZIMPOL-3 instrument attached to the Gregory-Coudé telescope of the IRSOL. The excellent performance of the optical systems during the observation ensured the high polarimetric sensitivity needed in this investigation.

A fascinating aspect of our study is the detection of a myriad of Stokes profiles in the analyzed prominence, which manifests the changing conditions along each slit and at different limb distances. In general, the observed Stokes profiles are broad and show multiple line features, such as a clear differentiation between the blue and red components of the line, different number of lobes in Stokes $V$, and conspicuous lineshifts. We assumed one- and two-component models to infer the physical properties of the prominence with the HAZEL code. In particular, the detected line features are usually better described using two components.

Our results reveal differences in some parameters depending on the limb distance. Close to the solar limb, both components contribute analogously to the output result given by HAZEL and have similar trends in some parameters. Specifically, they show LOS velocities of 1--3~km~s$\mathrm{^{-1}}$ and mostly horizontal magnetic fields of $\sim$30~G, while the thermal broadening diverges significantly between them ($\sim$10~km~s$\mathrm{^{-1}}$). On the other hand, further away from the limb, the filling factors inferred for each component diverge substantially, so we can distinguish a major component that mainly describes the properties of the observed prominence. Thus, at limb distances larger than 5\arcsec, the prominence usually shows LOS velocities below 3~km~s$\mathrm{^{-1}}$ and rather horizontal magnetic fields of 20--80~G, with the strongest fields being close to the center of the spine. Nonetheless, despite its small contribution, the minor component plays an essential role as it eases a good fit quality between the observed and synthetic profiles thanks to its large thermal broadening (10--15~km~s$\mathrm{^{-1}}$).

The analyzed prominence also shows intriguing imprints that are likely compatible with particular events. We found supersonic blueshifts in the prominence spine and a possible twist close to the foot. Moreover, these specific locations appear to be related to magnetic fields whose inclination angle departs from horizontal. These appealing features undoubtedly deserve a thorough investigation; however, we would also need a time series of spectropolarimetric data acquired at higher spatial resolution.  

Previous studies on prominences mainly focused on analyzing quiescent structures, so our results are not fully comparable to them. Regarding AR filaments, some investigations have attempted to interpret the physical properties retrieved from observational data \citepads[e.g.,][]{2009A&A...501.1113K, 2011A&A...526A..42S, 2012A&A...539A.131K, 2019A&A...625A.129D}. However, although filaments and prominences are counterparts of the same solar structure, we did not find magnetic fields of 600--700~G as were inferred by \citetads{2009A&A...501.1113K} and \citetads{2012A&A...539A.131K}, who only assumed the Zeeman effect to explain their observations in the \ion{He}{i} 10830~\AA\,triplet. According to \citetads{2019A&A...625A.129D}, the inference of such strong fields in AR filaments may be related to the fact that a significant part of the measured radiation comes from the underlying AR. In that case, the magnetic field given by a simple inversion strategy, such as considering the Milne-Eddington approximation or assuming a one-component atmosphere, may be overestimated in an AR filament. Instead, assuming a two-component atmosphere in an AR prominence has allowed us to infer a model atmosphere with weaker magnetic fields that can fit the observed profiles. Therefore, it is not straightforward to compare their properties one to one. Finally, after inspecting results retrieved in other off-limb structures, we note that \citetads{2005ESASP.596E..82R} reported magnetic fields of 50--60~G in spicules near an AR that are comparable to the ones we inferred.

Nevertheless, we should pay attention to some details in this study. First, the HAZEL code assumes that the incoming illumination to the target structure is due to the underlying photospheric quiet Sun. In this work, we analyzed an AR prominence, which means that the underlying illumination has different anisotropy than that considered, in principle, by the inversion code. Moreover, the incident radiation field in AR prominences may not be axially symmetric around the local vertical due to the dark and bright structures present below in the photosphere. This non-magnetic cause of symmetry breaking may have a significant impact on the scattering polarization signals, which unfortunately is not contemplated by HAZEL. Furthermore, we should consider the proximity of the prominence to the AR loops and the fact that the observations were acquired off-limb. Since we cannot resolve the information from the prominence and that from the AR loops, we expect a mixing of the signals coming from both structures. We mentioned this issue when describing the results on the field inclination, but it can also affect other parameters.
 
Finally, our results represent one of the plausible solutions that can be found. The main reason is that the parameters describing an observed feature are connected in many ways and, consequently, different models can explain a particular set of Stokes profiles. The degeneracy of the problem is thus too high to find a unique answer, as has also been noted in other studies \citepads[e.g.,][]{2019A&A...625A.129D}. After several attempts, we opted for using the inversion strategy that led to best-fit profiles similar to those observed and homogeneous output results in most of the slit positions. Other solutions, such as combining two atmospheric components of different LOS velocities or more complex scenarios, cannot be discarded as they might also account for the observed Stokes profiles. 

There are observing strategies that may help improve the reliability of the retrieved physical quantities in prominences; for instance, taking advantage of simultaneous spectropolarimetric observations in both the \ion{He}{i} 10830~\AA\,and D$\mathrm{_{3}}$ multiplets \citepads[as in][]{2009ApJ...703..114C}. Furthermore, analyzing full-Stokes measurements acquired at high spatial resolution and polarimetric sensitivity with next-generation large-aperture telescopes, such as DKIST \citepads{2020SoPh..295..172R} and EST \citepads{2022A&A...666A..21Q}, will undoubtedbly provide new insights into the physics of solar prominences.

\begin{acknowledgements}
We would like to thank the anonymous referee for helpful comments. We acknowledge the funding received from the European Research Council (ERC) under the European Union's Horizon 2020 research and innovation programme (Advanced grant agreement No 742265) and from the Agencia Estatal de Investigación del Ministerio de Ciencia, Innovación y Universidades (MCIU/AEI) under grant ``Polarimetric Inference of Magnetic Fields'' and the European Regional Development Fund (ERDF) with reference
PID2022-136563NB-I00/10.13039/501100011033. AAR acknowledges support from the Agencia Estatal de Investigaci\'on del
Ministerio de Ciencia, Innovaci\'on y Universidades (MCIU/AEI)
and the European Regional Development Fund (ERDF) through project PID2022-136563NB-I0. F.Z. acknowledges funding from the European’s Horizon 2020 research and innovation programme under grant agreement no 824135 and the Swiss National Science foundation under grant no PZ00P2\_215963. R.R. and F.Z. acknowledge funding from the Swiss National Science Foundation under grant number 200020\textbf{\textcolor{blue}{\_}}213147. IRSOL is supported by the Swiss Confederation (SEFRI), Canton Ticino, the city of Locarno and the local municipalities. This research has made use of NASA's Astrophysical Data System. The SDO data used in this work are courtesy of NASA/SDO and the AIA, EVE, and HMI science teams. We acknowledge the community effort devoted to the development of the following open-source packages that we used in this work: \texttt{numpy} \citepads{harris2020array} and \texttt{matplotlib} \citep{hunter:2007}. This work made use of \texttt{Astropy}: a community-developed core Python package and an ecosystem of tools and resources for astronomy \citepads{2022ApJ...935..167A}.
\end{acknowledgements}

%

\begin{thebibliography}{64}
\expandafter\ifx\csname natexlab\endcsname\relax\def\natexlab#1{#1}\fi

\bibitem[{{Antiochos} {et~al.}(1994){Antiochos}, {Dahlburg}, \&
  {Klimchuk}}]{1994ApJ...420L..41A}
{Antiochos}, S.~K., {Dahlburg}, R.~B., \& {Klimchuk}, J.~A. 1994, \apjl, 420,
  L41

\bibitem[{{Anzer} \& {Heinzel}(1999)}]{1999A&A...349..974A}
{Anzer}, U. \& {Heinzel}, P. 1999, \aap, 349, 974

\bibitem[{{Asensio Ramos} {et~al.}(2008){Asensio Ramos}, {Trujillo Bueno}, \&
  {Landi Degl'Innocenti}}]{2008ApJ...683..542A}
{Asensio Ramos}, A., {Trujillo Bueno}, J., \& {Landi Degl'Innocenti}, E. 2008,
  \apj, 683, 542

\bibitem[{{Astropy Collaboration} {et~al.}(2022){Astropy Collaboration},
  {Price-Whelan}, {Lim}, {Earl}, {Starkman}, {Bradley}, {Shupe}, {Patil},
  {Corrales}, {Brasseur}, {N{\"o}the}, {Donath}, {Tollerud}, {Morris},
  {Ginsburg}, {Vaher}, {Weaver}, {Tocknell}, {Jamieson}, {van Kerkwijk},
  {Robitaille}, {Merry}, {Bachetti}, {G{\"u}nther}, {Aldcroft},
  {Alvarado-Montes}, {Archibald}, {B{\'o}di}, {Bapat}, {Barentsen},
  {Baz{\'a}n}, {Biswas}, {Boquien}, {Burke}, {Cara}, {Cara}, {Conroy},
  {Conseil}, {Craig}, {Cross}, {Cruz}, {D'Eugenio}, {Dencheva}, {Devillepoix},
  {Dietrich}, {Eigenbrot}, {Erben}, {Ferreira}, {Foreman-Mackey}, {Fox},
  {Freij}, {Garg}, {Geda}, {Glattly}, {Gondhalekar}, {Gordon}, {Grant},
  {Greenfield}, {Groener}, {Guest}, {Gurovich}, {Handberg}, {Hart},
  {Hatfield-Dodds}, {Homeier}, {Hosseinzadeh}, {Jenness}, {Jones}, {Joseph},
  {Kalmbach}, {Karamehmetoglu}, {Ka{\l}uszy{\'n}ski}, {Kelley}, {Kern},
  {Kerzendorf}, {Koch}, {Kulumani}, {Lee}, {Ly}, {Ma}, {MacBride}, {Maljaars},
  {Muna}, {Murphy}, {Norman}, {O'Steen}, {Oman}, {Pacifici}, {Pascual},
  {Pascual-Granado}, {Patil}, {Perren}, {Pickering}, {Rastogi}, {Roulston},
  {Ryan}, {Rykoff}, {Sabater}, {Sakurikar}, {Salgado}, {Sanghi}, {Saunders},
  {Savchenko}, {Schwardt}, {Seifert-Eckert}, {Shih}, {Jain}, {Shukla}, {Sick},
  {Simpson}, {Singanamalla}, {Singer}, {Singhal}, {Sinha}, {Sip{\H{o}}cz},
  {Spitler}, {Stansby}, {Streicher}, {{\v{S}}umak}, {Swinbank}, {Taranu},
  {Tewary}, {Tremblay}, {de Val-Borro}, {Van Kooten}, {Vasovi{\'c}}, {Verma},
  {de Miranda Cardoso}, {Williams}, {Wilson}, {Winkel}, {Wood-Vasey}, {Xue},
  {Yoachim}, {Zhang}, {Zonca}, \& {Astropy Project
  Contributors}}]{2022ApJ...935..167A}
{Astropy Collaboration}, {Price-Whelan}, A.~M., {Lim}, P.~L., {et~al.} 2022,
  \apj, 935, 167

\bibitem[{{Bianda} {et~al.}(2009){Bianda}, {Ramelli}, \&
  {Gisler}}]{2009ASPC..405...17B}
{Bianda}, M., {Ramelli}, R., \& {Gisler}, D. 2009, in Astronomical Society of
  the Pacific Conference Series, Vol. 405, Solar Polarization 5: In Honor of
  Jan Stenflo, ed. S.~V. {Berdyugina}, K.~N. {Nagendra}, \& R.~{Ramelli}, 17

\bibitem[{{Casini} {et~al.}(2009){Casini}, {L{\'o}pez Ariste}, {Paletou}, \&
  {L{\'e}ger}}]{2009ApJ...703..114C}
{Casini}, R., {L{\'o}pez Ariste}, A., {Paletou}, F., \& {L{\'e}ger}, L. 2009,
  \apj, 703, 114

\bibitem[{{Casini} {et~al.}(2003){Casini}, {L{\'o}pez Ariste}, {Tomczyk}, \&
  {Lites}}]{2003ApJ...598L..67C}
{Casini}, R., {L{\'o}pez Ariste}, A., {Tomczyk}, S., \& {Lites}, B.~W. 2003,
  \apjl, 598, L67

\bibitem[{{Centeno} {et~al.}(2008){Centeno}, {Trujillo Bueno}, {Uitenbroek}, \&
  {Collados}}]{2008ApJ...677..742C}
{Centeno}, R., {Trujillo Bueno}, J., {Uitenbroek}, H., \& {Collados}, M. 2008,
  \apj, 677, 742

\bibitem[{{Collier} {et~al.}(2023){Collier}, {Hayes}, {Battaglia}, {Harra}, \&
  {Krucker}}]{2023A&A...671A..79C}
{Collier}, H., {Hayes}, L.~A., {Battaglia}, A.~F., {Harra}, L.~K., \&
  {Krucker}, S. 2023, \aap, 671, A79

\bibitem[{{Di Campli} {et~al.}(2020){Di Campli}, {Ramelli}, {Bianda}, {Furno},
  {Kumar Dhara}, \& {Belluzzi}}]{2020A&A...644A..89D}
{Di Campli}, R., {Ramelli}, R., {Bianda}, M., {et~al.} 2020, \aap, 644, A89

\bibitem[{{D{\'\i}az Baso} {et~al.}(2019){D{\'\i}az Baso}, {Mart{\'\i}nez
  Gonz{\'a}lez}, \& {Asensio Ramos}}]{2019A&A...625A.129D}
{D{\'\i}az Baso}, C.~J., {Mart{\'\i}nez Gonz{\'a}lez}, M.~J., \& {Asensio
  Ramos}, A. 2019, \aap, 625, A129

\bibitem[{{Felipe} {et~al.}(2017){Felipe}, {Collados}, {Khomenko}, {Rajaguru},
  {Franz}, {Kuckein}, \& {Asensio Ramos}}]{2017A&A...608A..97F}
{Felipe}, T., {Collados}, M., {Khomenko}, E., {et~al.} 2017, \aap, 608, A97

\bibitem[{{Gandorfer} \& {Povel}(1997)}]{1997A&A...328..381G}
{Gandorfer}, A.~M. \& {Povel}, H.~P. 1997, \aap, 328, 381

\bibitem[{{Gun{\'a}r} {et~al.}(2023){Gun{\'a}r}, {Labrosse}, {Luna},
  {Schmieder}, {Heinzel}, {Kucera}, {Levens}, {L{\'o}pez Ariste}, {Mackay}, \&
  {Zapi{\'o}r}}]{2023SSRv..219...33G}
{Gun{\'a}r}, S., {Labrosse}, N., {Luna}, M., {et~al.} 2023, \ssr, 219, 33

\bibitem[{Harris {et~al.}(2020)Harris, Millman, van~der Walt, Gommers,
  Virtanen, Cournapeau, Wieser, Taylor, Berg, Smith, Kern, Picus, Hoyer, van
  Kerkwijk, Brett, Haldane, del R{\'{i}}o, Wiebe, Peterson,
  G{\'{e}}rard-Marchant, Sheppard, Reddy, Weckesser, Abbasi, Gohlke, \&
  Oliphant}]{harris2020array}
Harris, C.~R., Millman, K.~J., van~der Walt, S.~J., {et~al.} 2020, Nature, 585,
  357

\bibitem[{Hunter(2007)}]{hunter:2007}
Hunter, J.~D. 2007, Computing in Science \& Engineering, 9, 90

\bibitem[{{Kalewicz} \& {Bommier}(2019)}]{2019A&A...629A.138K}
{Kalewicz}, T. \& {Bommier}, V. 2019, \aap, 629, A138

\bibitem[{{Kemp} {et~al.}(1984){Kemp}, {Macek}, \&
  {Nehring}}]{1984ApJ...278..863K}
{Kemp}, J.~C., {Macek}, J.~H., \& {Nehring}, F.~W. 1984, \apj, 278, 863

\bibitem[{{Kippenhahn} \& {Schl{\"u}ter}(1957)}]{1957ZA.....43...36K}
{Kippenhahn}, R. \& {Schl{\"u}ter}, A. 1957, \zap, 43, 36

\bibitem[{{Kucera}(2015)}]{2015ASSL..415...79K}
{Kucera}, T.~A. 2015, in Astrophysics and Space Science Library, Vol. 415,
  Solar Prominences, ed. J.-C. {Vial} \& O.~{Engvold}, 79

\bibitem[{{Kuckein} {et~al.}(2009){Kuckein}, {Centeno}, {Mart{\'\i}nez Pillet},
  {Casini}, {Manso Sainz}, \& {Shimizu}}]{2009A&A...501.1113K}
{Kuckein}, C., {Centeno}, R., {Mart{\'\i}nez Pillet}, V., {et~al.} 2009, \aap,
  501, 1113

\bibitem[{{Kuckein} {et~al.}(2012){Kuckein}, {Mart{\'\i}nez Pillet}, \&
  {Centeno}}]{2012A&A...539A.131K}
{Kuckein}, C., {Mart{\'\i}nez Pillet}, V., \& {Centeno}, R. 2012, \aap, 539,
  A131

\bibitem[{{Labrosse} \& {Gouttebroze}(2004)}]{2004ApJ...617..614L}
{Labrosse}, N. \& {Gouttebroze}, P. 2004, \apj, 617, 614

\bibitem[{{Labrosse} {et~al.}(2010){Labrosse}, {Heinzel}, {Vial}, {Kucera},
  {Parenti}, {Gun{\'a}r}, {Schmieder}, \& {Kilper}}]{2010SSRv..151..243L}
{Labrosse}, N., {Heinzel}, P., {Vial}, J.~C., {et~al.} 2010, \ssr, 151, 243

\bibitem[{{Lemen} {et~al.}(2012){Lemen}, {Title}, {Akin}, {Boerner}, {Chou},
  {Drake}, {Duncan}, {Edwards}, {Friedlaender}, {Heyman}, {Hurlburt}, {Katz},
  {Kushner}, {Levay}, {Lindgren}, {Mathur}, {McFeaters}, {Mitchell}, {Rehse},
  {Schrijver}, {Springer}, {Stern}, {Tarbell}, {Wuelser}, {Wolfson}, {Yanari},
  {Bookbinder}, {Cheimets}, {Caldwell}, {Deluca}, {Gates}, {Golub}, {Park},
  {Podgorski}, {Bush}, {Scherrer}, {Gummin}, {Smith}, {Auker}, {Jerram},
  {Pool}, {Soufli}, {Windt}, {Beardsley}, {Clapp}, {Lang}, \&
  {Waltham}}]{2012SoPh..275...17L}
{Lemen}, J.~R., {Title}, A.~M., {Akin}, D.~J., {et~al.} 2012, \solphys, 275, 17

\bibitem[{{Leroy} {et~al.}(1983){Leroy}, {Bommier}, \&
  {Sahal-Brechot}}]{1983SoPh...83..135L}
{Leroy}, J.~L., {Bommier}, V., \& {Sahal-Brechot}, S. 1983, \solphys, 83, 135

\bibitem[{{Leroy} {et~al.}(1984){Leroy}, {Bommier}, \&
  {Sahal-Brechot}}]{1984A&A...131...33L}
{Leroy}, J.~L., {Bommier}, V., \& {Sahal-Brechot}, S. 1984, \aap, 131, 33

\bibitem[{{L{\'o}pez Ariste} \& {Aulanier}(2007)}]{2007ASPC..368..291L}
{L{\'o}pez Ariste}, A. \& {Aulanier}, G. 2007, in Astronomical Society of the
  Pacific Conference Series, Vol. 368, The Physics of Chromospheric Plasmas,
  ed. P.~{Heinzel}, I.~{Dorotovi{\v{c}}}, \& R.~J. {Rutten}, 291

\bibitem[{{L{\'o}pez Ariste} {et~al.}(2006){L{\'o}pez Ariste}, {Aulanier},
  {Schmieder}, \& {Sainz Dalda}}]{2006A&A...456..725L}
{L{\'o}pez Ariste}, A., {Aulanier}, G., {Schmieder}, B., \& {Sainz Dalda}, A.
  2006, \aap, 456, 725

\bibitem[{{L{\'o}pez Ariste} \& {Casini}(2003)}]{2003ApJ...582L..51L}
{L{\'o}pez Ariste}, A. \& {Casini}, R. 2003, \apjl, 582, L51

\bibitem[{{Mackay} {et~al.}(2010){Mackay}, {Karpen}, {Ballester}, {Schmieder},
  \& {Aulanier}}]{2010SSRv..151..333M}
{Mackay}, D.~H., {Karpen}, J.~T., {Ballester}, J.~L., {Schmieder}, B., \&
  {Aulanier}, G. 2010, \ssr, 151, 333

\bibitem[{{Mart{\'\i}nez Gonz{\'a}lez} {et~al.}(2015){Mart{\'\i}nez
  Gonz{\'a}lez}, {Manso Sainz}, {Asensio Ramos}, {Beck}, {de la Cruz
  Rodr{\'\i}guez}, \& {D{\'\i}az}}]{2015ApJ...802....3M}
{Mart{\'\i}nez Gonz{\'a}lez}, M.~J., {Manso Sainz}, R., {Asensio Ramos}, A.,
  {et~al.} 2015, \apj, 802, 3

\bibitem[{{Merenda} {et~al.}(2006){Merenda}, {Trujillo Bueno}, {Landi
  Degl'Innocenti}, \& {Collados}}]{2006ApJ...642..554M}
{Merenda}, L., {Trujillo Bueno}, J., {Landi Degl'Innocenti}, E., \& {Collados},
  M. 2006, \apj, 642, 554

\bibitem[{{Orozco Su{\'a}rez} {et~al.}(2012){Orozco Su{\'a}rez}, {Asensio
  Ramos}, \& {Trujillo Bueno}}]{2012ApJ...761L..25O}
{Orozco Su{\'a}rez}, D., {Asensio Ramos}, A., \& {Trujillo Bueno}, J. 2012,
  \apjl, 761, L25

\bibitem[{{Orozco Su{\'a}rez} {et~al.}(2014){Orozco Su{\'a}rez}, {Asensio
  Ramos}, \& {Trujillo Bueno}}]{2014A&A...566A..46O}
{Orozco Su{\'a}rez}, D., {Asensio Ramos}, A., \& {Trujillo Bueno}, J. 2014,
  \aap, 566, A46

\bibitem[{{Osherovich}(1989)}]{1989ApJ...336.1041O}
{Osherovich}, V.~A. 1989, \apj, 336, 1041

\bibitem[{{Paletou} {et~al.}(2001){Paletou}, {L{\'o}pez Ariste}, {Bommier}, \&
  {Semel}}]{2001A&A...375L..39P}
{Paletou}, F., {L{\'o}pez Ariste}, A., {Bommier}, V., \& {Semel}, M. 2001,
  \aap, 375, L39

\bibitem[{{Parenti}(2014)}]{2014LRSP...11....1P}
{Parenti}, S. 2014, Living Reviews in Solar Physics, 11, 1

\bibitem[{{Peat}(2023)}]{2023PhDT.........1P}
{Peat}, A.~W. 2023, PhD thesis, University of Glasgow, UK

\bibitem[{{P{\'e}cseli} \& {Engvold}(2000)}]{2000SoPh..194...73P}
{P{\'e}cseli}, H. \& {Engvold}, O. 2000, \solphys, 194, 73

\bibitem[{{Pesnell} {et~al.}(2012){Pesnell}, {Thompson}, \&
  {Chamberlin}}]{2012SoPh..275....3P}
{Pesnell}, W.~D., {Thompson}, B.~J., \& {Chamberlin}, P.~C. 2012, \solphys,
  275, 3

\bibitem[{{Pettit}(1932)}]{1932ApJ....76....9P}
{Pettit}, E. 1932, \apj, 76, 9

\bibitem[{Pierce(2000)}]{Pierce2000}
Pierce, K. 2000, in Allen's Astrophysical Quantities, 4th edn., ed. A.~N. Cox
  (New York: Springer)

\bibitem[{{Quintero Noda} {et~al.}(2022){Quintero Noda}, {Schlichenmaier},
  {Bellot Rubio}, {L{\"o}fdahl}, {Khomenko}, {Jur{\v{c}}{\'a}k}, {Leenaarts},
  {Kuckein}, {Gonz{\'a}lez Manrique}, {Gun{\'a}r}, {Nelson}, {de la Cruz
  Rodr{\'\i}guez}, {Tziotziou}, {Tsiropoula}, {Aulanier}, {Aboudarham},
  {Allegri}, {Alsina Ballester}, {Amans}, {Asensio Ramos}, {Bail{\'e}n},
  {Balaguer}, {Baldini}, {Balthasar}, {Barata}, {Barczynski}, {Barreto
  Cabrera}, {Baur}, {B{\'e}chet}, {Beck}, {Bel{\'\i}o-As{\'\i}n},
  {Bello-Gonz{\'a}lez}, {Belluzzi}, {Bentley}, {Berdyugina}, {Berghmans},
  {Berlicki}, {Berrilli}, {Berkefeld}, {Bettonvil}, {Bianda}, {Bienes
  P{\'e}rez}, {Bonaque-Gonz{\'a}lez}, {Braj{\v{s}}a}, {Bommier}, {Bourdin},
  {Burgos Mart{\'\i}n}, {Calchetti}, {Calcines}, {Calvo Tovar}, {Campbell},
  {Carballo-Mart{\'\i}n}, {Carbone}, {Carlin}, {Carlsson}, {Castro L{\'o}pez},
  {Cavaller}, {Cavallini}, {Cauzzi}, {Cecconi}, {Chulani}, {Cirami},
  {Consolini}, {Coretti}, {Cosentino}, {C{\'o}zar-Castellano}, {Dalmasse},
  {Danilovic}, {De Juan Ovelar}, {Del Moro}, {del Pino Alem{\'a}n}, {del Toro
  Iniesta}, {Denker}, {Dhara}, {Di Marcantonio}, {D{\'\i}az Baso}, {Diercke},
  {Dineva}, {D{\'\i}az-Garc{\'\i}a}, {Doerr}, {Doyle}, {Erdelyi}, {Ermolli},
  {Escobar Rodr{\'\i}guez}, {Esteban Pozuelo}, {Faurobert}, {Felipe}, {Feller},
  {Feijoo Amoedo}, {Femen{\'\i}a Castell{\'a}}, {Fernandes}, {Ferro
  Rodr{\'\i}guez}, {Figueroa}, {Fletcher}, {Franco Ordovas}, {Gafeira},
  {Gardenghi}, {Gelly}, {Giorgi}, {Gisler}, {Giovannelli}, {Gonz{\'a}lez},
  {Gonz{\'a}lez}, {Gonz{\'a}lez-Cava}, {Gonz{\'a}lez Garc{\'\i}a},
  {G{\"o}m{\"o}ry}, {Gracia}, {Grauf}, {Greco}, {Grivel}, {Guerreiro},
  {Guglielmino}, {Hammerschlag}, {Hanslmeier}, {Hansteen}, {Heinzel},
  {Hern{\'a}ndez-Delgado}, {Hern{\'a}ndez Su{\'a}rez}, {Hidalgo}, {Hill},
  {Hizberger}, {Hofmeister}, {J{\"a}gers}, {Janett}, {Jarolim}, {Jess},
  {Jim{\'e}nez Mej{\'\i}as}, {Jolissaint}, {Kamlah}, {Kapit{\'a}n},
  {Ka{\v{s}}parov{\'a}}, {Keller}, {Kentischer}, {Kiselman}, {Kleint},
  {Klvana}, {Kontogiannis}, {Krishnappa}, {Ku{\v{c}}era}, {Labrosse}, {Lagg},
  {Landi Degl'Innocenti}, {Langlois}, {Lafon}, {Laforgue}, {Le Men}, {Lepori},
  {Lepreti}, {Lindberg}, {Lilje}, {L{\'o}pez Ariste}, {L{\'o}pez
  Fern{\'a}ndez}, {L{\'o}pez Jim{\'e}nez}, {L{\'o}pez L{\'o}pez}, {Manso
  Sainz}, {Marassi}, {Marco de la Rosa}, {Marino}, {Marrero}, {Mart{\'\i}n},
  {Mart{\'\i}n G{\'a}lvez}, {Mart{\'\i}n Hernando}, {Masciadri}, {Mart{\'\i}nez
  Gonz{\'a}lez}, {Matta-G{\'o}mez}, {Mato}, {Mathioudakis}, {Matthews}, {Mein},
  {Merlos Garc{\'\i}a}, {Moity}, {Montilla}, {Molinaro}, {Molodij}, {Montoya},
  {Munari}, {Murabito}, {N{\'u}{\~n}ez Cagigal}, {Oliviero}, {Orozco
  Su{\'a}rez}, {Ortiz}, {Padilla-Hern{\'a}ndez}, {Pa{\'e}z Ma{\~n}{\'a}},
  {Paletou}, {Pancorbo}, {Pastor Ca{\~n}edo}, {Pastor Yabar}, {Peat},
  {Pedichini}, {Peixinho}, {Pe{\~n}ate}, {P{\'e}rez de Taoro}, {Peter},
  {Petrovay}, {Piazzesi}, {Pietropaolo}, {Pleier}, {Poedts}, {P{\"o}tzi},
  {Podladchikova}, {Prieto}, {Quintero Nehrkorn}, {Ramelli}, {Ramos Sapena},
  {Rasilla}, {Reardon}, {Rebolo}, {Regalado Olivares}, {Reyes
  Garc{\'\i}a-Talavera}, {Riethm{\"u}ller}, {Rimmele}, {Rodr{\'\i}guez
  Delgado}, {Rodr{\'\i}guez Gonz{\'a}lez}, {Rodr{\'\i}guez-Losada},
  {Rodr{\'\i}guez Ramos}, {Romano}, {Roth}, {Rouppe van der Voort}, {Rudawy},
  {Ruiz de Galarreta}, {Ryb{\'a}k}, {Salvade}, {S{\'a}nchez-Capuchino},
  {S{\'a}nchez Rodr{\'\i}guez}, {Sangiorgi}, {Say{\`e}de}, {Scharmer},
  {Scheiffelen}, {Schmidt}, {Schmieder}, {Scir{\`e}}, {Scuderi}, {Siegel},
  {Sigwarth}, {Sim{\~o}es}, {Snik}, {Sliepen}, {Sobotka}, {Socas-Navarro},
  {Sola La Serna}, {Solanki}, {Soler Trujillo}, {Soltau}, {Sordini}, {Sosa
  M{\'e}ndez}, {Stangalini}, {Steiner}, {Stenflo}, {{\v{S}}t{\v{e}}p{\'a}n},
  {Strassmeier}, {Sudar}, {Suematsu}, {S{\"u}tterlin}, {Tallon}, {Temmer},
  {Tenegi}, {Tritschler}, {Trujillo Bueno}, {Turchi}, {Utz}, {van Harten}, {van
  Noort}, {van Werkhoven}, {Vansintjan}, {Vaz Cedillo}, {Vega Reyes}, {Verma},
  {Veronig}, {Viavattene}, {Vitas}, {V{\"o}gler}, {von der L{\"u}he},
  {Volkmer}, {Waldmann}, {Walton}, {Wisniewska}, {Zeman}, {Zeuner}, {Zhang},
  {Zuccarello}, \& {Collados}}]{2022A&A...666A..21Q}
{Quintero Noda}, C., {Schlichenmaier}, R., {Bellot Rubio}, L.~R., {et~al.}
  2022, \aap, 666, A21

\bibitem[{{Ramelli} {et~al.}(2010){Ramelli}, {Balemi}, {Bianda}, {Defilippis},
  {Gamma}, {Hagenbuch}, {Rogantini}, {Steiner}, \&
  {Stenflo}}]{2010SPIE.7735E..1YR}
{Ramelli}, R., {Balemi}, S., {Bianda}, M., {et~al.} 2010, in Society of
  Photo-Optical Instrumentation Engineers (SPIE) Conference Series, Vol. 7735,
  Ground-based and Airborne Instrumentation for Astronomy III, ed. I.~S.
  {McLean}, S.~K. {Ramsay}, \& H.~{Takami}, 77351Y

\bibitem[{{Ramelli} \& {Bianda}(2005)}]{2005ASSL..320..215R}
{Ramelli}, R. \& {Bianda}, M. 2005, in Astrophysics and Space Science Library,
  Vol. 320, Solar Magnetic Phenomena, ed. A.~{Hanslmeier}, A.~{Veronig}, \&
  M.~{Messerotti}, 215--218

\bibitem[{{Ramelli} {et~al.}(2005){Ramelli}, {Bianda}, {Trujillo Bueno},
  {Merenda}, \& {Stenflo}}]{2005ESASP.596E..82R}
{Ramelli}, R., {Bianda}, M., {Trujillo Bueno}, J., {Merenda}, L., \& {Stenflo},
  J.~O. 2005, in ESA Special Publication, Vol. 596, Chromospheric and Coronal
  Magnetic Fields, ed. D.~E. {Innes}, A.~{Lagg}, \& S.~A. {Solanki}, 82.1

\bibitem[{{Ramelli} {et~al.}(2012){Ramelli}, {Stellmacher}, {Wiehr}, \&
  {Bianda}}]{2012SoPh..281..697R}
{Ramelli}, R., {Stellmacher}, G., {Wiehr}, E., \& {Bianda}, M. 2012, \solphys,
  281, 697

\bibitem[{{Ramelli} {et~al.}(2011){Ramelli}, {Trujillo Bueno}, {Bianda}, \&
  {Asensio Ramos}}]{2011ASPC..437..109R}
{Ramelli}, R., {Trujillo Bueno}, J., {Bianda}, M., \& {Asensio Ramos}, A. 2011,
  in Astronomical Society of the Pacific Conference Series, Vol. 437, Solar
  Polarization 6, ed. J.~R. {Kuhn}, D.~M. {Harrington}, H.~{Lin}, S.~V.
  {Berdyugina}, J.~{Trujillo-Bueno}, S.~L. {Keil}, \& T.~{Rimmele}, 109

\bibitem[{{Rimmele} {et~al.}(2020){Rimmele}, {Warner}, {Keil}, {Goode},
  {Kn{\"o}lker}, {Kuhn}, {Rosner}, {McMullin}, {Casini}, {Lin}, {W{\"o}ger},
  {von der L{\"u}he}, {Tritschler}, {Davey}, {de Wijn}, {Elmore}, {Fehlmann},
  {Harrington}, {Jaeggli}, {Rast}, {Schad}, {Schmidt}, {Mathioudakis},
  {Mickey}, {Anan}, {Beck}, {Marshall}, {Jeffers}, {Oschmann}, {Beard},
  {Berst}, {Cowan}, {Craig}, {Cross}, {Cummings}, {Donnelly}, {de Vanssay},
  {Eigenbrot}, {Ferayorni}, {Foster}, {Galapon}, {Gedrites}, {Gonzales},
  {Goodrich}, {Gregory}, {Guzman}, {Guzzo}, {Hegwer}, {Hubbard}, {Hubbard},
  {Johansson}, {Johnson}, {Liang}, {Liang}, {McQuillen}, {Mayer}, {Newman},
  {Onodera}, {Phelps}, {Puentes}, {Richards}, {Rimmele}, {Sekulic}, {Shimko},
  {Simison}, {Smith}, {Starman}, {Sueoka}, {Summers}, {Szabo}, {Szabo},
  {Wampler}, {Williams}, \& {White}}]{2020SoPh..295..172R}
{Rimmele}, T.~R., {Warner}, M., {Keil}, S.~L., {et~al.} 2020, \solphys, 295,
  172

\bibitem[{{Sasso} {et~al.}(2011){Sasso}, {Lagg}, \&
  {Solanki}}]{2011A&A...526A..42S}
{Sasso}, C., {Lagg}, A., \& {Solanki}, S.~K. 2011, \aap, 526, A42

\bibitem[{{Schmieder} {et~al.}(2010){Schmieder}, {Chandra}, {Berlicki}, \&
  {Mein}}]{2010A&A...514A..68S}
{Schmieder}, B., {Chandra}, R., {Berlicki}, A., \& {Mein}, P. 2010, \aap, 514,
  A68

\bibitem[{{Schmieder} {et~al.}(2015){Schmieder}, {L{\'o}pez Ariste}, {Levens},
  {Labrosse}, \& {Dalmasse}}]{2015IAUS..305..275S}
{Schmieder}, B., {L{\'o}pez Ariste}, A., {Levens}, P., {Labrosse}, N., \&
  {Dalmasse}, K. 2015, in IAU Symposium, Vol. 305, Polarimetry, ed. K.~N.
  {Nagendra}, S.~{Bagnulo}, R.~{Centeno}, \& M.~{Jes{\'u}s Mart{\'\i}nez
  Gonz{\'a}lez}, 275--281

\bibitem[{{Schwarz}(1978)}]{1978AnSta...6..461S}
{Schwarz}, G. 1978, Annals of Statistics, 6, 461

\bibitem[{{Tandberg-Hanssen}(1974)}]{1974GAM....12.....T}
{Tandberg-Hanssen}, E. 1974, {Solar Prominences}, Vol.~12

\bibitem[{{Tandberg-Hanssen}(1995)}]{1995ASSL..199.....T}
{Tandberg-Hanssen}, E. 1995, {The nature of solar prominences}, Vol. 199

\bibitem[{{Trujillo Bueno} \& {Asensio Ramos}(2007)}]{2007ApJ...655..642T}
{Trujillo Bueno}, J. \& {Asensio Ramos}, A. 2007, \apj, 655, 642

\bibitem[{{Trujillo Bueno} \& {del Pino
  Alem{\'a}n}(2022)}]{2022ARA&A..60..415T}
{Trujillo Bueno}, J. \& {del Pino Alem{\'a}n}, T. 2022, \araa, 60, 415

\bibitem[{{Trujillo Bueno} {et~al.}(2002){Trujillo Bueno}, {Landi
  Degl'Innocenti}, {Collados}, {Merenda}, \& {Manso
  Sainz}}]{2002Natur.415..403T}
{Trujillo Bueno}, J., {Landi Degl'Innocenti}, E., {Collados}, M., {Merenda},
  L., \& {Manso Sainz}, R. 2002, \nat, 415, 403

\bibitem[{{Trujillo Bueno} {et~al.}(2005){Trujillo Bueno}, {Merenda},
  {Centeno}, {Collados}, \& {Landi Degl'Innocenti}}]{2005ApJ...619L.191T}
{Trujillo Bueno}, J., {Merenda}, L., {Centeno}, R., {Collados}, M., \& {Landi
  Degl'Innocenti}, E. 2005, \apjl, 619, L191

\bibitem[{{Vial} \& {Engvold}(2015)}]{2015ASSL..415.....V}
{Vial}, J.-C. \& {Engvold}, O., eds. 2015, Astrophysics and Space Science
  Library, Vol. 415, {Solar Prominences}

\bibitem[{{Zeuner} {et~al.}(2022){Zeuner}, {Gisler}, {Bianda}, {Ramelli}, \&
  {Berdyugina}}]{2022SPIE12184E..0TZ}
{Zeuner}, F., {Gisler}, D., {Bianda}, M., {Ramelli}, R., \& {Berdyugina}, S.~V.
  2022, in Society of Photo-Optical Instrumentation Engineers (SPIE) Conference
  Series, Vol. 12184, Ground-based and Airborne Instrumentation for Astronomy
  IX, ed. C.~J. {Evans}, J.~J. {Bryant}, \& K.~{Motohara}, 121840T

\bibitem[{{Zirin}(1975)}]{1975ApJ...199L..63Z}
{Zirin}, H. 1975, \apjl, 199, L63

\bibitem[{{Zirker} {et~al.}(1998){Zirker}, {Engvold}, \&
  {Martin}}]{1998Natur.396..440Z}
{Zirker}, J.~B., {Engvold}, O., \& {Martin}, S.~F. 1998, \nat, 396, 440

\end{thebibliography}

%


\end{document}